\documentclass[journal]{IEEEtran}

\usepackage{cite}
\usepackage{amsmath,amssymb,amsfonts}
\usepackage{algorithmic}
\usepackage{graphicx}
\usepackage{textcomp}

\usepackage[USenglish]{babel}
\usepackage{blindtext}
\usepackage{tabularray}
\usepackage{booktabs}

\usepackage{placeins}

\usepackage{bm}
\usepackage{balance}
\usepackage{subfigure}
\usepackage{textcomp}
\usepackage{xcolor}
\usepackage{float} % https://stackoverflow.com/questions/1673942/latex-table-positioning
\usepackage{graphicx}
\usepackage{url}

\usepackage{diagbox}
\usepackage{algorithmic}
\usepackage{verbatim}
\usepackage{multirow}
\usepackage{balance}
\usepackage{array}
\usepackage{tabu}
\usepackage{longtable}[=v4.13]
\usepackage{booktabs} % for borders and merged ranges
\usepackage{subfigure}
\usepackage{caption}
\usepackage{enumitem}
\usepackage{tabularx}

\usepackage{comment}
\usepackage{url}
\usepackage{listings}
\definecolor{codegreen}{rgb}{0,0.6,0}
\definecolor{codegray}{rgb}{0.5,0.5,0.5}
\definecolor{codepurple}{rgb}{0.58,0,0.82}
\definecolor{backcolour}{rgb}{0.95,0.95,0.92}
\lstdefinestyle{mystyle}{
    backgroundcolor=\color{backcolour},   
    commentstyle=\color{codegreen},
    keywordstyle=\color{magenta},
    numberstyle=\tiny\color{codegray},
    stringstyle=\color{codepurple},
    basicstyle=\small,
    breakatwhitespace=false,         
    breaklines=true,                 
    captionpos=b,                    
    keepspaces=true,                 
    numbers=left,                    
    numbersep=5pt,                  
    showspaces=false,                
    showstringspaces=false,
    showtabs=false,                  
    tabsize=2
}
\usepackage{soul} % prashanth
\makeatletter
\AtBeginDocument{\DeclareMathVersion{bold}
\SetSymbolFont{operators}{bold}{T1}{times}{b}{n}
\DeclareSymbolFont{NewLetters}{T1}{times}{m}{it}
\SetSymbolFont{NewLetters}{bold}{T1}{times}{b}{it}
\SetMathAlphabet{\mathrm}{bold}{T1}{times}{b}{n}
\SetMathAlphabet{\mathit}{bold}{T1}{times}{b}{it}
\SetMathAlphabet{\mathbf}{bold}{T1}{times}{b}{n}
\SetMathAlphabet{\mathtt}{bold}{OT1}{pcr}{b}{n}
\SetSymbolFont{symbols}{bold}{OMS}{cmsy}{b}{n}
\renewcommand\boldmath{\@nomath\boldmath\mathversion{bold}}}
\makeatother

\def\BibTeX{{\rm B\kern-.05em{\sc i\kern-.025em b}\kern-.08em
    T\kern-.1667em\lower.7ex\hbox{E}\kern-.125emX}}
    
\begin{document}
%\history{Date of publication xxxx 00, 0000, date of current version xxxx 00, 0000.}
%\doi{10.1109/ACCESS.2026.xxxx}

\title{Precision-Aware Variable Bit Processing Elements for Hardware-Efficient Systolic Array Designs}

% \author{
% \IEEEauthorblockN{Dantu Nandini Devi}
% \IEEEauthorblockA{IIIT-Bangalore\\
% \textit{Bangalore, India}\\
% dantunandini.devi@iiitb.ac.in
% \orcidlink{0009-0006-8233-5120}
% }
% \and
% \IEEEauthorblockN{Madhav Rao}
% \IEEEauthorblockA{IIIT-Bangalore,  \\
% \textit{Bangalore, India}\\
% mr@iiitb.ac.in
% \orcidlink{0000-0003-2278-9148}}
% }
\author{
    \uppercase{Dantu Nandini Devi} \authorrefmark{1}, %,
    %\uppercase{Kanish R} \authorrefmark{1},
    and Madhav Rao  \authorrefmark{1} \IEEEmembership{Senior Member, IEEE}
    %\IEEEmembership{Member, IEEE}
}

\author{Dantu Nandini Devi,
    Madhav Rao,
\thanks{Dantu Nandini Devi was with the Department
of Electronics \& Communication Engineering, IIIT-Bangalore,
Karnataka, India }% <-this % stops a space
\thanks{Madhav Rao is a faculty working at ECE Department of IIIT-Bangalore.}% <-this % stops a space
}

%\address[1]{International Institute of %Information Technology Bangalore, KA 560100, %India}

%\author{Dantu Nandini Devi\orcidlink{0009-0006-%8233-5120},Madhav Rao\orcidlink{0000-0003-2278-%9148}\\{\%{dantunandini.devi,mr\}}@iiitb.ac.in\\Internation%al Institute of Information Technology, %Bangalore}

        % <-this % stops a space
%\thanks{This paper was produced by the IEEE Publication Technology Group. They are in Piscataway, NJ.}% <-this % stops a space
%\thanks{Manuscript received April 19, 2021; revised August 16, 2021.}

%\markboth{IEEE TRANSACTIONS ON VLSI SYSTEMS,~Vol.~XX, No.~X, %Sept~2025}%
%{Shell \MakeLowercase{\textit{et al.}}: A Sample Article %Using IEEEtran.cls for IEEE Journals}


%\markboth
%{Dantu Nandini Devi \headeretal: Precision-%Aware Systolic Array Designs}
%{Dantu Nandini Devi \headeretal: Precision-%Aware  Systolic Array Designs}
% example, ``This work was supported in part by the U.S. Department of
% Commerce under Grant BS123456.''}

%\corresp{Corresponding author: Madhav Rao (e-%mail: mr@iiitb.ac.in)}

\maketitle

%%%To do:
%% Add few % metric for CNN in the Abstract & %% Conclusion -- Done !!
%% Add more recent Citations & remove few of our %%% own old citations & less important ones --- Done !!

%%% Notation representing the SA and PEs inside them from the table... Done !!

%%% One diagram of SA with varied bit-width PEs.

%%% Finalize Title... Done !!

%% Ask Nandini to add website for putting all plots & data.

\begin{abstract}
Systolic arrays~(SAs) have emerged as prominent hardware accelerators for matrix operations in deep learning, while floating point number formats enable precision control across computational domains. Approximate computing involves trading precision for improved performance in applications where exact accuracy is not essential. This research investigates approximate computing techniques for floating-point~(FP) multipliers in Weight Stationary Systolic Arrays, focusing on IEEE 754~(FP32), TensorFloat-32~(TF32), and Brain Floating point~(BF16) formats. By integrating partial-product-matrix~(PPM) column truncation with positive and negative compressors in the FP multiplier architecture, we optimize the trade-off between computational efficiency and accuracy. NSGA-II optimization algorithm was employed to explore the vast design space for evolving FP multiplier designs, towards achieving substantial hardware improvements while maintaining 
acceptable output quality.
%acceptable output quality. 
Substantial hardware benefits were observed in the FP multiplier designs across various applications, while preserving output quality.
%Evaluations across various applications, presents
%%we have observed 
%remarkable hardware benefits while maintaining high %output quality. 
The FP approximated Processing Elements designed in the
SA was found to offer comparable CNN accuracy for models trained on MNIST, F-MNIST, and CIFAR-10 dataset.
The FP approximated SA designs that fall 
in the top 10 CNN performance
offered substantial 
hardware gains in the range of 
%82\% to 92\% 
 66\% - 92\% footprint savings,  
 %80\% to 93\% 
 60\% - 93\% of power benefits with 
 21\% - 54\%
 %27\% to 60\% 
 improvement in the delay when compared with the corresponding exact implementations mentioned in the literature for running the model trained on CIFAR-10 dataset. The TF32 and BF16 approximated SA designs also achieved substantial gains while maintaining comparable CNN accuracy.
Our findings confirm that targeted approximation in FP multiplier design significantly improves the efficiency of hardware accelerators for error-tolerant applications, establishing an effective approach to hardware resource optimization in contemporary computing architectures. Furthermore, the framework and hardware design files are made available for further usage to the designers' and researchers community.
\end{abstract}


\section{Introduction}

Approximate computing has emerged as a powerful paradigm that offers substantial advantages in error-tolerant contexts, particularly within image and signal processing applications~\cite{apprx-fp1, apprx-fp2,
am1, am2, am3, yashaswi-adder, new1, bindu-error-diluted, new2, bindu-isqed24, new3, nandini-sa-isqed24}. This approach strategically trades computational precision for significant improvements in performance, energy efficiency, and hardware footprint. The benefits of approximate computing extend across both software domains, through algorithmic optimizations~\cite{algo1, algo4, hw-sw}, and hardware implementations, where it effectively reduces power consumption while minimizing circuit footprint and shortening critical paths~\cite{apprx1, apprx2, apprx3, hardware-savings1}. As computational demands continue to grow exponentially while hardware scaling faces fundamental physical limitations, approximate computing's ability to navigate these constraints has attracted substantial interest within the VLSI design community. Its capacity to achieve an optimal balance between performance, energy efficiency, and accuracy, positions it as a compelling solution for addressing the ever demanding  computational requirements in increasingly resource-constrained environments.

Artificial Intelligence (AI) and Machine Learning (ML) workloads have demonstrated remarkable performance acceleration when executed on dedicated co-processors~\cite{nvdla1, nvdla2, google-tpu1, google-tpu2}. At the core of modern AI systems, Convolutional Neural Networks (CNNs) and Fully Connected Networks (FCNs) are computationally dominated by multiplier and accumulator (MAC) operations~\cite{mult-cnn1, mult-cnn3, mult-cnn11, mult-cnn12, mult-cnn13, mult-cnn14, mult-cnn15, mult-cnn16}. While these neural network architectures incorporate a diverse range of functional components, many are fundamentally implemented using MAC units as their computational building blocks. Recent research has extensively investigated the integration of approximate processing elements (PEs) within CNNs, systematically evaluating the complex trade-offs between hardware efficiency gains and potential accuracy degradation~\cite{alwann, bindu-raghava-iccd, balanced-mac, cnn-apprx-mult1, cnn-apprx-mult2, cnn-apprx-mult3, cnn-apprx-mult4}. Contemporary research efforts have increasingly focused on optimizing multiplier architectures and usage patterns to minimize accuracy losses while maximizing hardware benefits, with some approaches exploring the possibility of tuning network parameters during training phases to compensate for approximation-induced errors~\cite{apprx-training1, apprx-training2}.

Systolic Arrays (SAs) represent specialized hardware accelerators meticulously optimized for the efficient execution of General Matrix Multiplication (GEMM) and Convolution operations~\cite{pruning, new-sa3, SAOld, sa-new1, new-sa-power-efficient, sa-new2, sa3}. These arrays are architecturally composed of a two-dimensional (2D) arrangement of Processing Elements (PEs), with each PE containing a Multiply-Accumulate (MAC) unit and associated storage registers. SAs are typically classified into three distinct configurations based on their data flow patterns: Input Stationary (IS), Weight Stationary (WS), and Output Stationary (OS), with each configuration offering optimized performance characteristics for specific computational requirements and workloads. Within each PE, the multiplier components of the MAC units constitute the most hardware-intensive and resource-demanding elements. 
By strategically introducing approximations in these multipliers, significant hardware efficiency gains are  realized, albeit with carefully managed trade-offs in computational accuracy. This approach effectively leverages the benefits of reduced circuit complexity and resource utilization while maintaining acceptable output quality.
Previously in~\cite{bindu-raghava-iccd}, it was empirically shown that localized positive or negative multipliers for individual layers in CNNs
offers much better model performance.
%which is attributed to the error dilution on usage of 
%different multipliers
Hence, compensating for errors within each kernel channels by employing inexact multipliers in a SA hardware accelerator is aimed towards generating close to exact CNN outputs. Indeed, it was found later that retrieved configurations are very homogeneous across PEs of SA, with small local deviations which is expected from the heuristic nature of the algorithm adopted in this work.

Previous research efforts exploring the integration of approximate multipliers within SAs have predominantly focused on implementing and evaluating single types of approximate multipliers~\cite{SA-factored-hybrid, sa-cnn1, sa-cnn2}. Studies that have incorporated multiple approximate multiplier architectures have largely been limited to specific image processing applications using predetermined kernel weights on restricted datasets, or constrained to integer multiplier implementations~\cite{apprx-image1, apprx-image2, apprx-image3, new1,
%nandini_opst, karthik_paper, 
nandini-sa-isqed24}. Our study significantly expands upon this foundation by specifically focusing on diverse floating-point representation formats, including the industry-standard FP32, as well as the more specialized TF32 and BF16 formats, and systematically approximating their corresponding multiplier implementations. 
This comprehensive approach aims to provide a more nuanced and thorough understanding of the potential benefits and trade-offs inherent in applying approximate computing principles to SAs across a broader spectrum of applications and data representation formats. 
Considering these data formats are standardized and well established, pareto-front SA design solutions 
offer the most balanced structure in terms of hardware benefits and accuracy control.
The authors have made the designs freely available in~\cite{website} for further usage to the researchers and designers community.

\section{Proposed Design}

\begin{figure}
    \centering
    \includegraphics[width=0.6\linewidth]{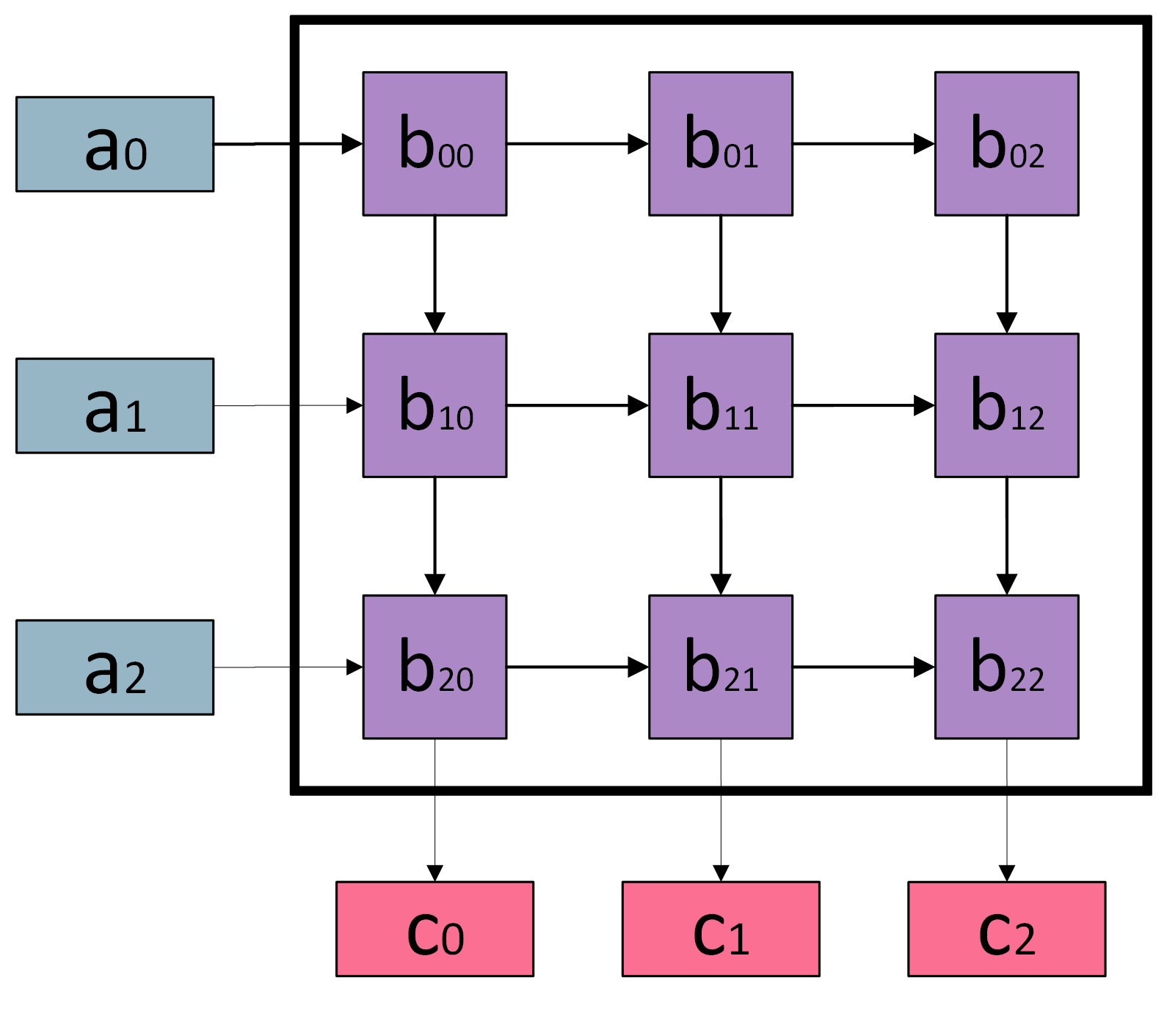}
    \caption{Schematic showing the input data flow and kernel weights in the Weight Stationary SA architecture.}
    \label{fig:sawtst}
    % \vspace{-6mm}
\end{figure}

\begin{figure}
    \centering
    \includegraphics[width=1\linewidth]{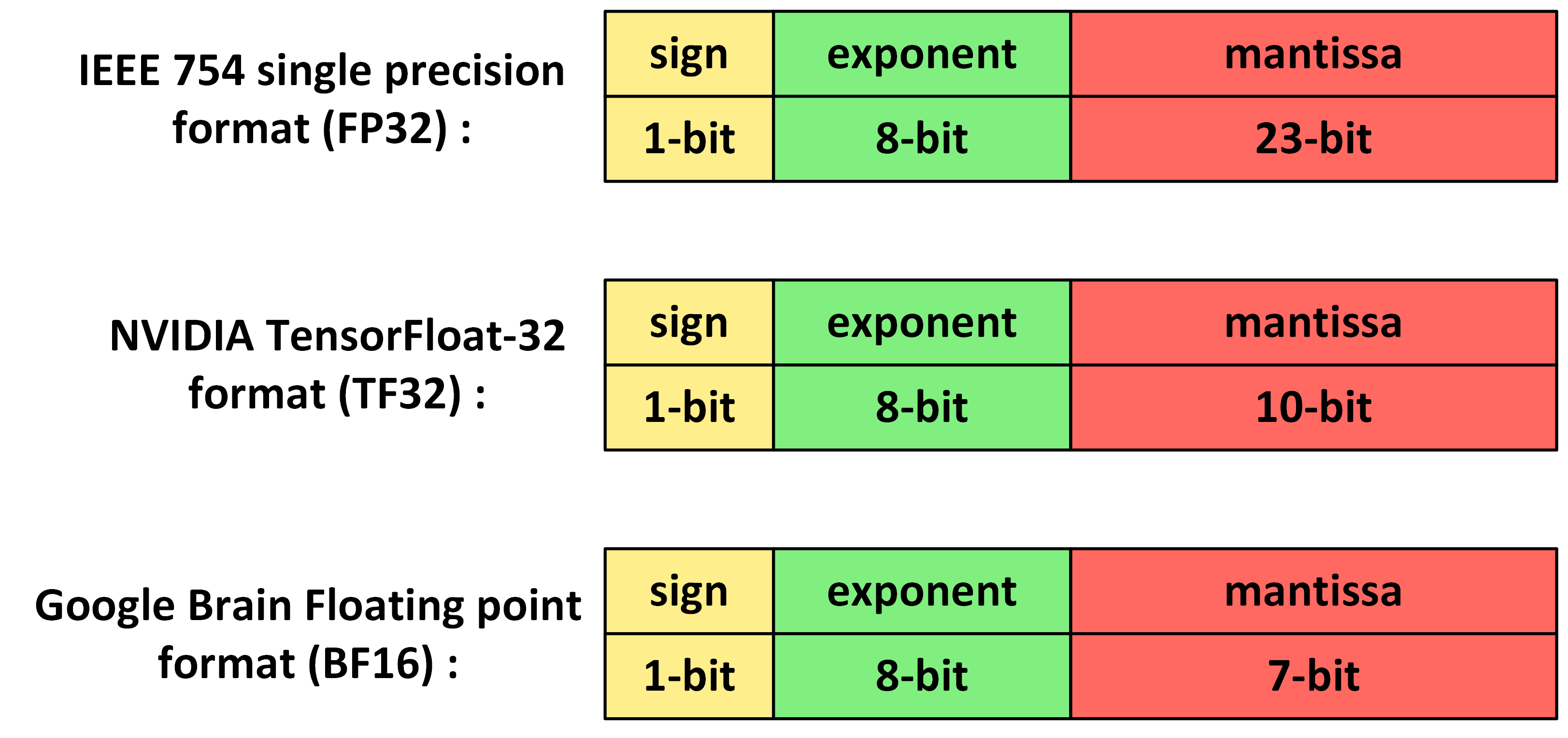}
    \caption{Schematic showing the three floating-point data formats with Sign, Exponent, and Mantissa bits.}
    \label{fig:fp_formats}
    % \vspace{-6mm}
\end{figure}

FP representation %is a critical method 
is typically employed for computing 
%for approximating 
real numbers across diverse numerical ranges. This research investigates the optimization of FP multipliers within WS SAs, focusing on achieving an optimal balance between computational efficiency and numerical accuracy. 
Figure~\ref{fig:sawtst} shows the input data flow from left to right, while output traverses from top to bottom in a weight stationary SA. 
Floating-point representation, characterized by its sign, exponent, and mantissa components, enables wide-ranging numerical computations but introduces inherent complexity in multiplication operations.
Figure~\ref{fig:fp_formats} shows the three popular floating-point formats and its bit-wise representations. Note that while exponent bits for all the three depicted formats are same, the mantissa bit-width differs.

The proposed methodology introduces strategic approximations in the mantissa multiplication stage through three innovative techniques. First, we determine the optimal number of columns for truncation during partial product generation. Second, we identify the ideal bit count for post-truncation approximation. Third, we implement Positive and Negative compressors in the partial product reduction stage. These approaches generate multipliers with nuanced error distributions, which are strategically arranged to neutralize computational errors while preserving the advantages of hardware approximation. By concentrating on  WS SAs, the research addresses FP arithmetic challenges through a hardware-efficient framework. This study aims to advance approximate computing by refining the balance between computational speed and precision, enabling new applications in high-performance, energy-efficient computing that rely on floating-point operations.

\begin{figure*}
    \centering
    \includegraphics[width=1\linewidth]{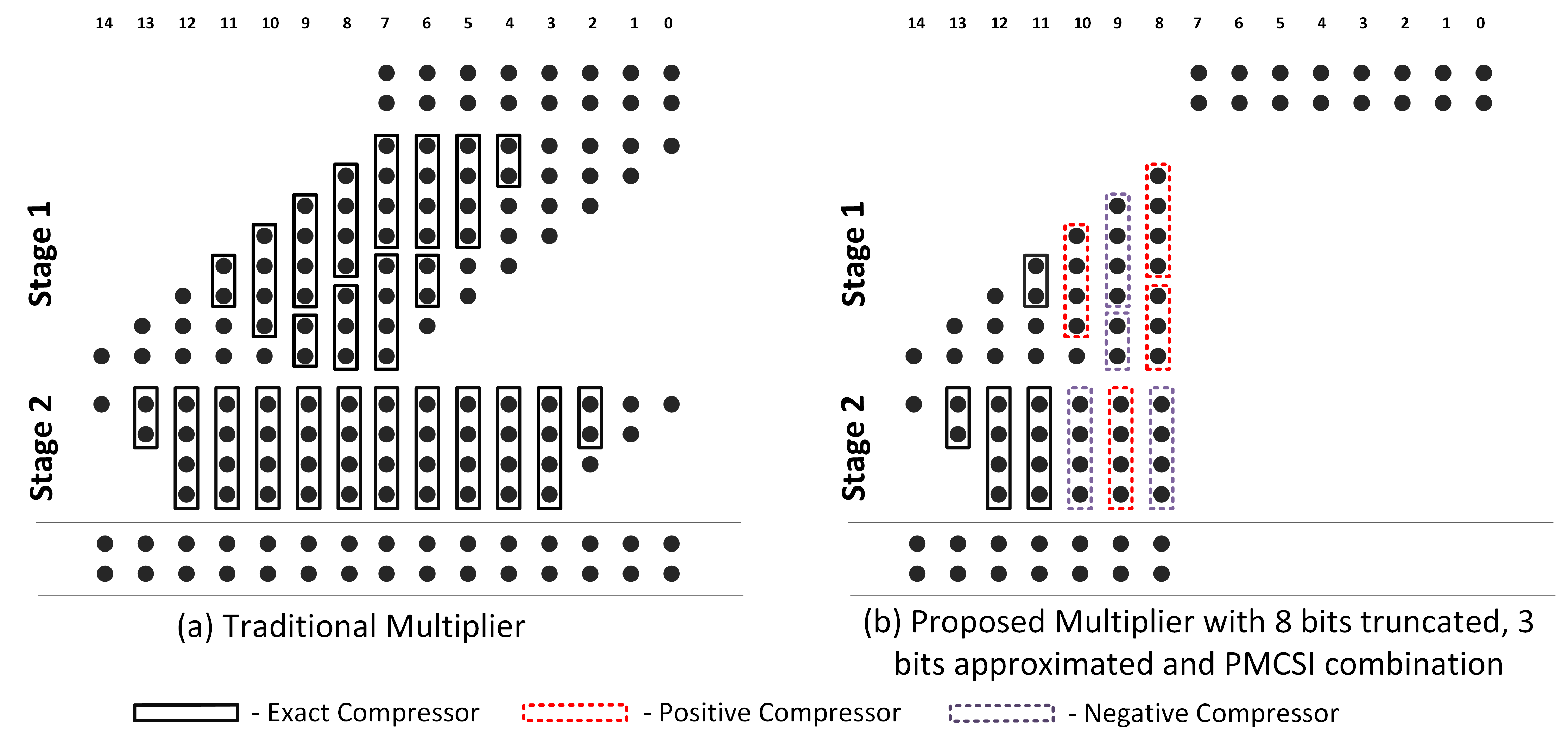}
    \caption{Example showing the difference between the reduction tree of (a) traditional multiplier and (b) proposed multiplier where the first 8 bits are truncated, 3 bits are approximated by employing Positive Column-Cum-Stage-wise Multiplier~(PMCSI) compressor configuration.  Note that the number of bits to be truncated and approximated is deduced from the Evolutionary Algorithm run.}
    %truncation and approximation bits are der}
    \label{fig:proposed_design_example}
    % \vspace{-6mm}
\end{figure*}
Our research synthesizes two pioneering approaches in approximate computing for multiplier design. We adapt the right-shift approximate multiplication algorithm, originally proposed for sequential in-memory computing~\cite{saketh-vlsid24}, to create novel combinational multiplier designs. This adaptation transforms the sequential method—which selectively computes the most significant product bits—into a parallel design suitable for diverse hardware architectures. Simultaneously, we integrate comprehensive compressor application strategies developed in prior work~\cite{bindu-error-diluted}, encompassing Non-Interleaving, Stage-Wise, Column-Wise, and Column-cum-Stage-Wise methodologies. Initially demonstrated on an 8$\times$8 bit unsigned multiplier, these strategies yield eight distinct multiplier configurations, each presenting unique error distribution characteristics. 
These are also classified as positive and negative error
distributed compressors which is attributed to error distribution being dominant either on positive or negative side.
%more error
The aim of the earlier work~\cite{bindu-error-diluted} was to combine these set of positive and negative compressors to dilute the errors which otherwise keeps elevating due
to the usage of only unidirectional inexact compressors, and also avoid using any additional error-compensation
circuits. 
We also incorporated exact compressors as one of the option in this compressors selection, hence overall 9 such options are available to be picked for designing PE for SA.
By integrating these methodological innovations, our work aims to develop FP multipliers that leverage approximate computing benefits while maintaining acceptable accuracy for error-tolerant computational scenarios. This holistic approach potentially broadens the applicability of approximate multiplication techniques across compute-intensive applications, offering unprecedented opportunities for optimizing hardware in error-tolerant computing environments.

\begin{comment}
Our research introduces a novel three-pronged strategy for approximation in the mantissa multiplication stage. We develop a systematic method for determining the optimal number of columns for truncation in partial product generation, directly impacting the efficiency-accuracy trade-off. We establish criteria for identifying the ideal number of bits for approximation post-truncation, ensuring minimal impact on computational accuracy. Additionally, we implement an innovative approach utilizing Positive and Negative compressors in the partial product reduction stage, creating a balanced error distributed system.
\end{comment}
Figure~\ref{fig:proposed_design_example} highlights the key distinctions between a traditional multiplier reduction tree and our proposed methodology. To illustrate our approach, we provide an example of an 8-bit multiplier reduction tree. Figure~\ref{fig:proposed_design_example}(a) represents the reduction tree of a conventional multiplier, utilizing exact compressors to achieve precise outputs. In contrast, Figure~\ref{fig:proposed_design_example}(b) demonstrates our proposed reduction tree, which integrates the core concepts of our evolutionary design strategy. In this configuration, the least significant 8 bits are truncated, and the subsequent 3 bits are approximated using a tailored combination of approximate compressors. This design adheres to the Positive Column-Cum-Stage-wise Multiplier~(PMCSI) framework, as outlined in~\cite{bindu-error-diluted}.

\section{Design Space Exploration}

\begin{figure}[!htp]
    \centering
    \includegraphics[width=0.75\linewidth]{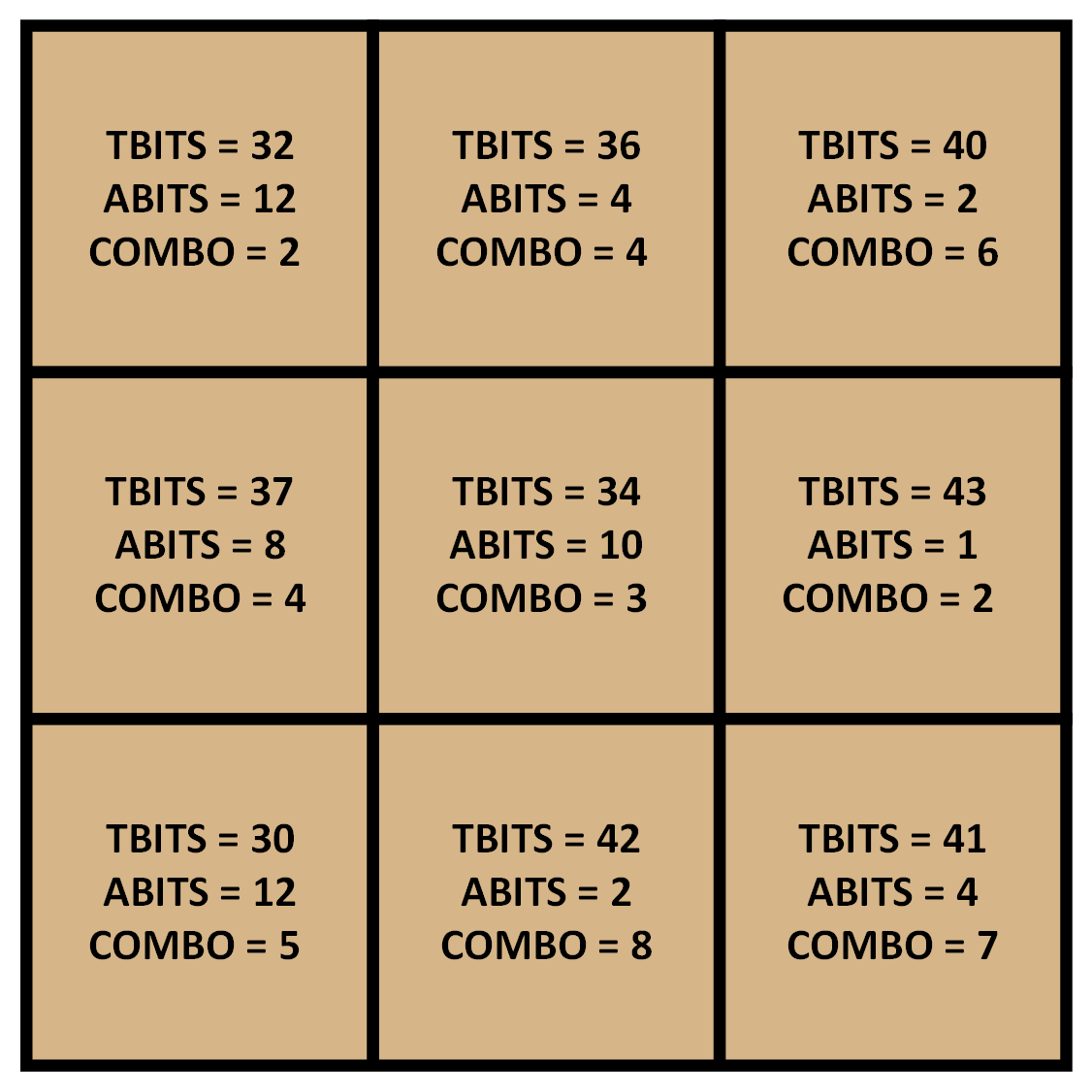}
    \caption{Example configuration evolved by NSGA-II for a $3 \times 3$ FP32-based SA, showing varied bit truncation, approximation strategies, and compressor configurations across individual PEs.}
    %truncation and approximation bits are der}
    \label{fig:sa_proposed_design_example}
    % \vspace{-6mm}
\end{figure}

The introduction of dynamic truncation and approximation bits significantly expands the solution space. For a 3$\times$3 SA with 9 PEs, each PE now has three independent variables to optimize: truncation bits~, approximation bits~, and compressor sequence. 
By combining the previously mentioned methods, we find $8\times n\times(n+1)\div2$ configurations for each PE, where $n$ represents the number of columns in the exact reduction stage. In a 3$\times$3 SA with 9 PEs we will have design space of $(8\times n\times(n+1)\div2)^9$.
%, vastly exceeding the previous space of %$(n\times(n+1)\div2)\times(8^9) $ combinations.
%This leads to a configuration space of possibilities, vastly exceeding the previous space of $COMBO^9$ %combinations. 
Additionally all these design needs to be evaluated on atleast four different parameters including three (footprint, latency, power) and output performance.
This expanded search space in multiple objective dimensions  necessitates a more viable strategy to implement.
Non-Dominated Sorting Genetic Algorithm II (NSGA-II) is a multi-objective evolutionary algorithm
that suits our requirement to
%that reduces computational complexity, eliminates the %need for specifying a sharing parameter, and applies a %non-elitism approach.
%NSGA-II algorithm
%implementation of the NSGA-II algorithm to 
efficiently navigate the increased complexity while maintaining practical optimization time-frames. Figure~\ref{fig:sa_proposed_design_example} illustrates an exemplary configuration generated by the NSGA-II evolutionary algorithm for a $3 \times 3$ FP32-based SA. In this optimized design, each Processing Element (PE) features a unique combination of bit truncation levels, approximation strategies, and compressor configurations. The algorithm strategically varies these parameters across different PEs within the array to achieve an optimal balance between computational accuracy and hardware efficiency metrics.

In our proposed methodology, each chromosome in the NSGA-II framework encodes a complete SA configuration. For every PE within the array, three design variables are considered: the number of truncated mantissa columns (TBITS), the number of bits approximated after truncation (ABITS), and the choice of compressor type (COMBO) drawn from eight available variants. Thus, for a 3$\times$3 SA consisting of nine PEs, each chromosome contains 27 genes. This representation ensures that the chromosome structure fully captures the approximation strategy for the entire array.

The evolutionary operators of NSGA-II are adapted to effectively explore this configuration space. Crossover is applied at the chromosome level, where one-point or two-point crossover exchanges contiguous segments of PEs between two parent solutions. This allows entire truncation/approximation/compressor sequences to be inherited, thereby preserving locally optimized structures. Mutation is applied at the gene level by randomly perturbing TBITS ($\pm$), altering ABITS, or changing the compressor type. These operators maintain genetic diversity within the population and prevent premature convergence, enabling the algorithm to balance exploration and exploitation of the design space.

The NSGA-II optimization framework takes as input the application-level quality metric, represented by SSIM or classification accuracy depending on the target application, along with key hardware evaluation metrics, including worst-case delay for performance assessment, hardware footprint for resource utilization, and power consumption of the corresponding systolic array (SA). Based on these multi-objective constraints and evaluation parameters, the framework generates an optimized SA configuration that represents a Pareto-optimal trade-off among computational accuracy, performance, area, and power efficiency.
For the encoding scheme used within the NSGA-II framework, each processing element (PE) configuration is represented using three design variables: the number of truncated mantissa columns (TBITS), the number of approximated bits following truncation (ABITS), and the compressor selection parameter (COMBO), which is chosen from eight available compressor variants. The complete chromosome encoding is constructed by concatenating the configurations of all required PEs, where the total number of PE configurations depends on the dimensions of the systolic array.

The compressor implementations are derived from prior work and include both exact and approximate variants~\cite{bindu-error-diluted}. Specifically, eight approximate compressor configurations are used, which are categorized into positive and negative error-distributed compressors depending on the dominant error direction. By mixing positive and negative compressors across PEs, NSGA-II evolves design solutions that balance accumulated errors without the need for additional error-compensation circuits. Exact compressors are also retained in the design pool, enabling hybrid solutions where approximation is selectively applied.

\begin{figure*}[!htp]
\centering
\resizebox{0.9\textwidth}{!}{
\setlength{\tabcolsep}{0.5pt} % Default value: 6pt
    \begin{tabular}{c c c}
     % \centering
     %\setlength{\tabcolsep}{1pt}
        \textbf{\Huge Averaging Filter : FP32} & \textbf{\Huge Averaging Filter : TF32} & \textbf{\Huge Averaging Filter : BF16} \\ 
    
        \includegraphics[scale=1]{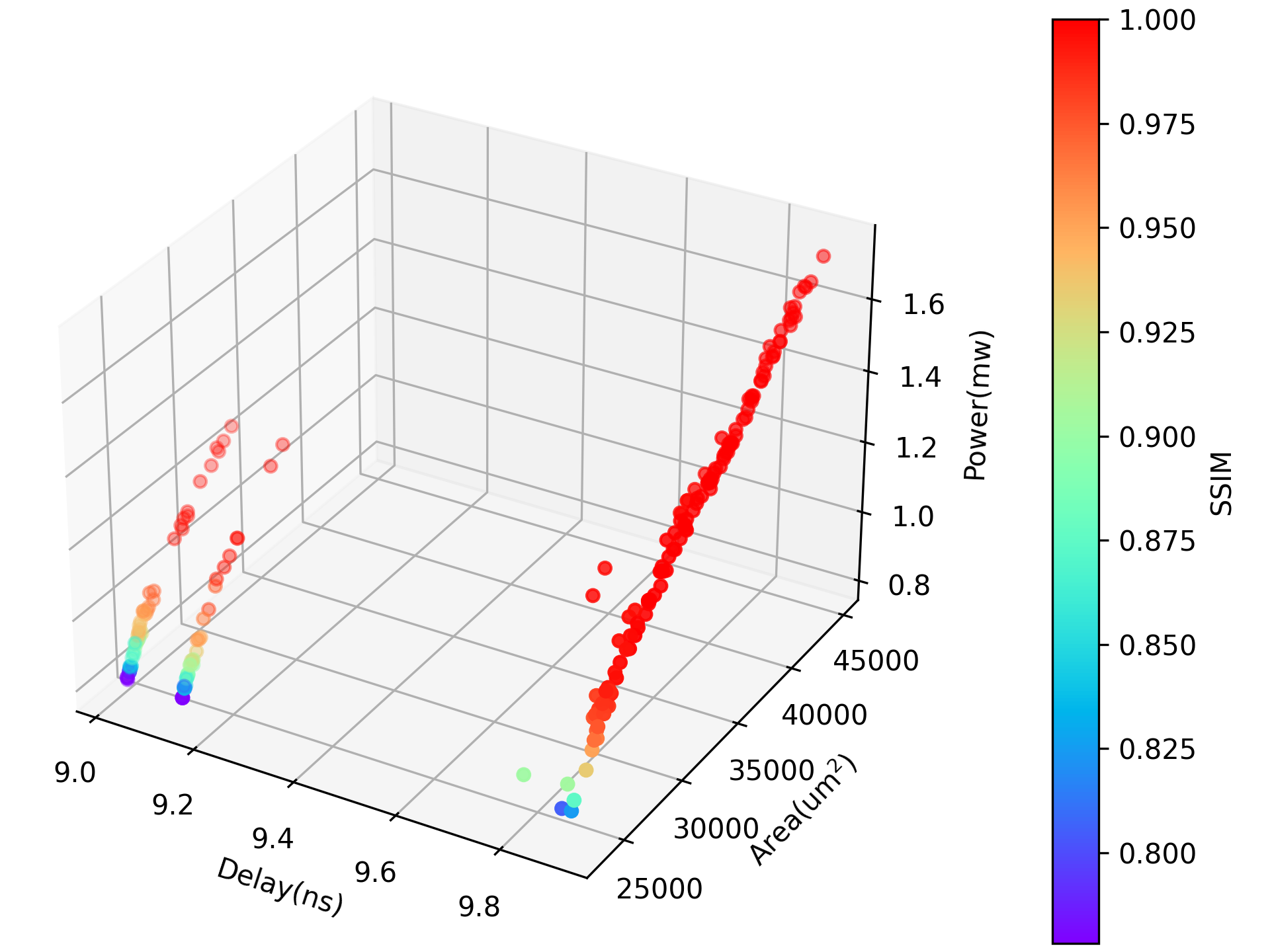} &
        \includegraphics[scale=1]{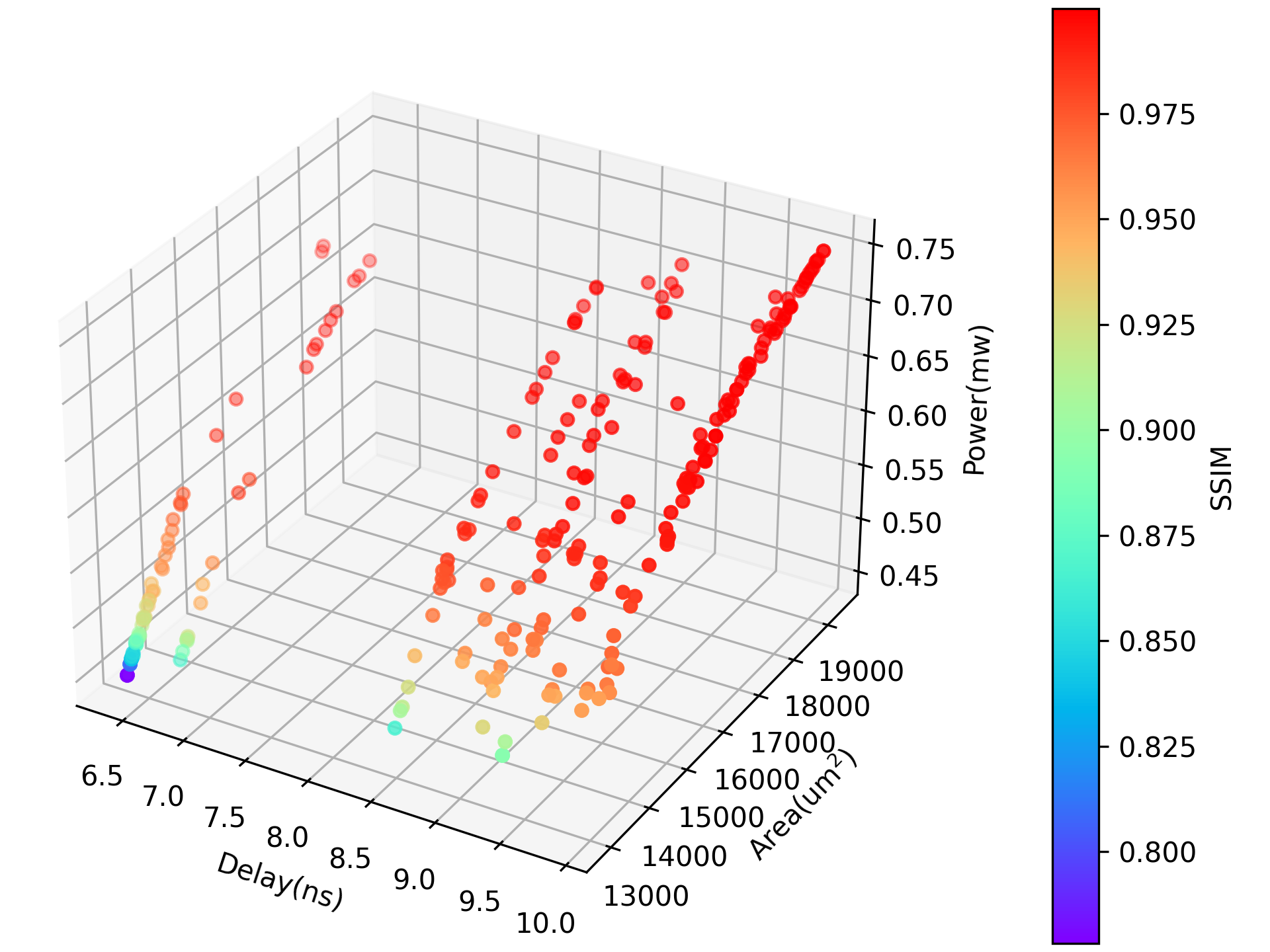} &
        \includegraphics[scale=1]{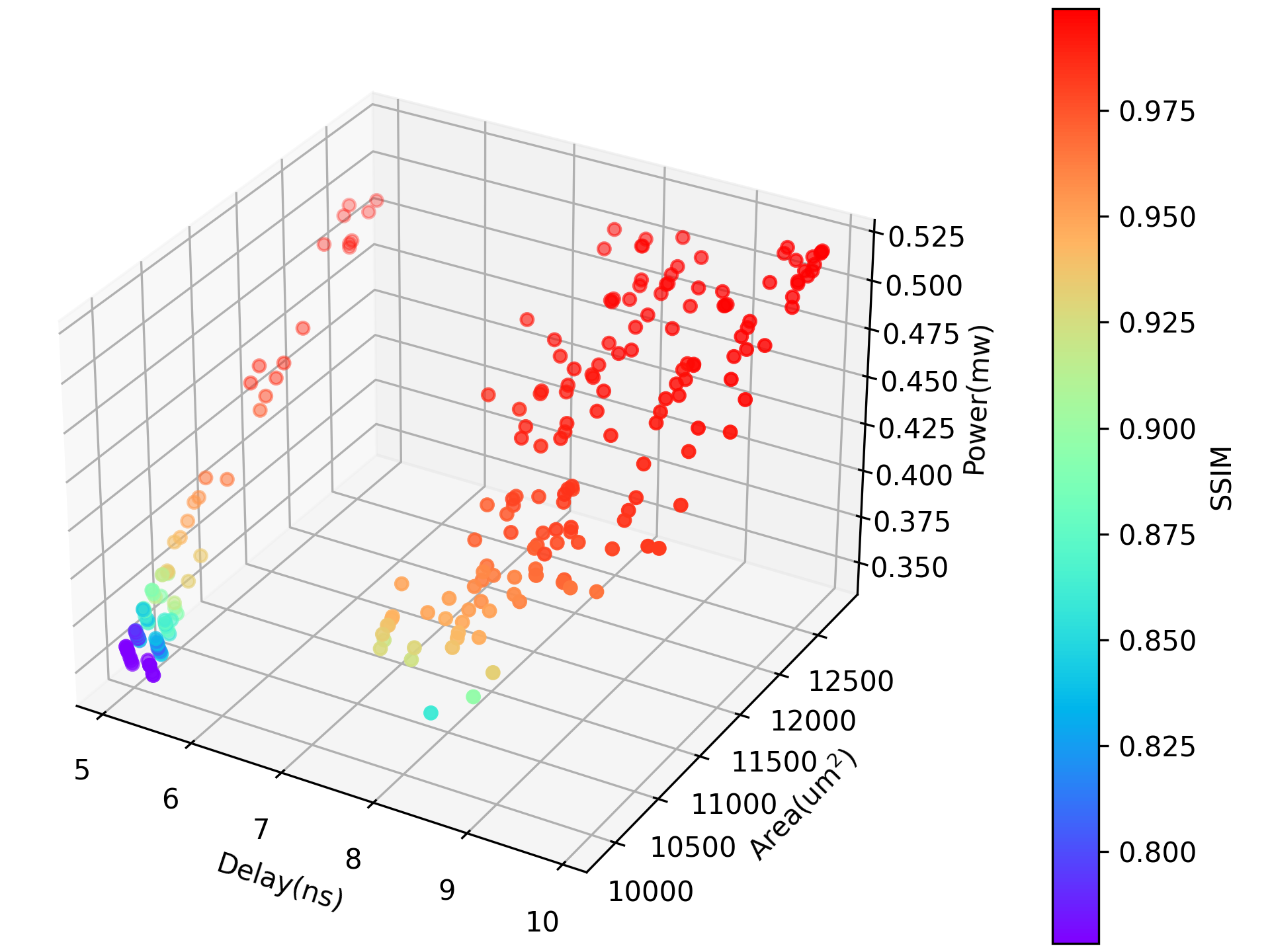}
        \\

         &  &  \\
         \textbf{\Huge Gaussian Filter : FP32} & \textbf{\Huge Gaussian Filter : TF32} & \textbf{\Huge Gaussian Filter : BF16} \\ 
    
        \includegraphics[scale=1]{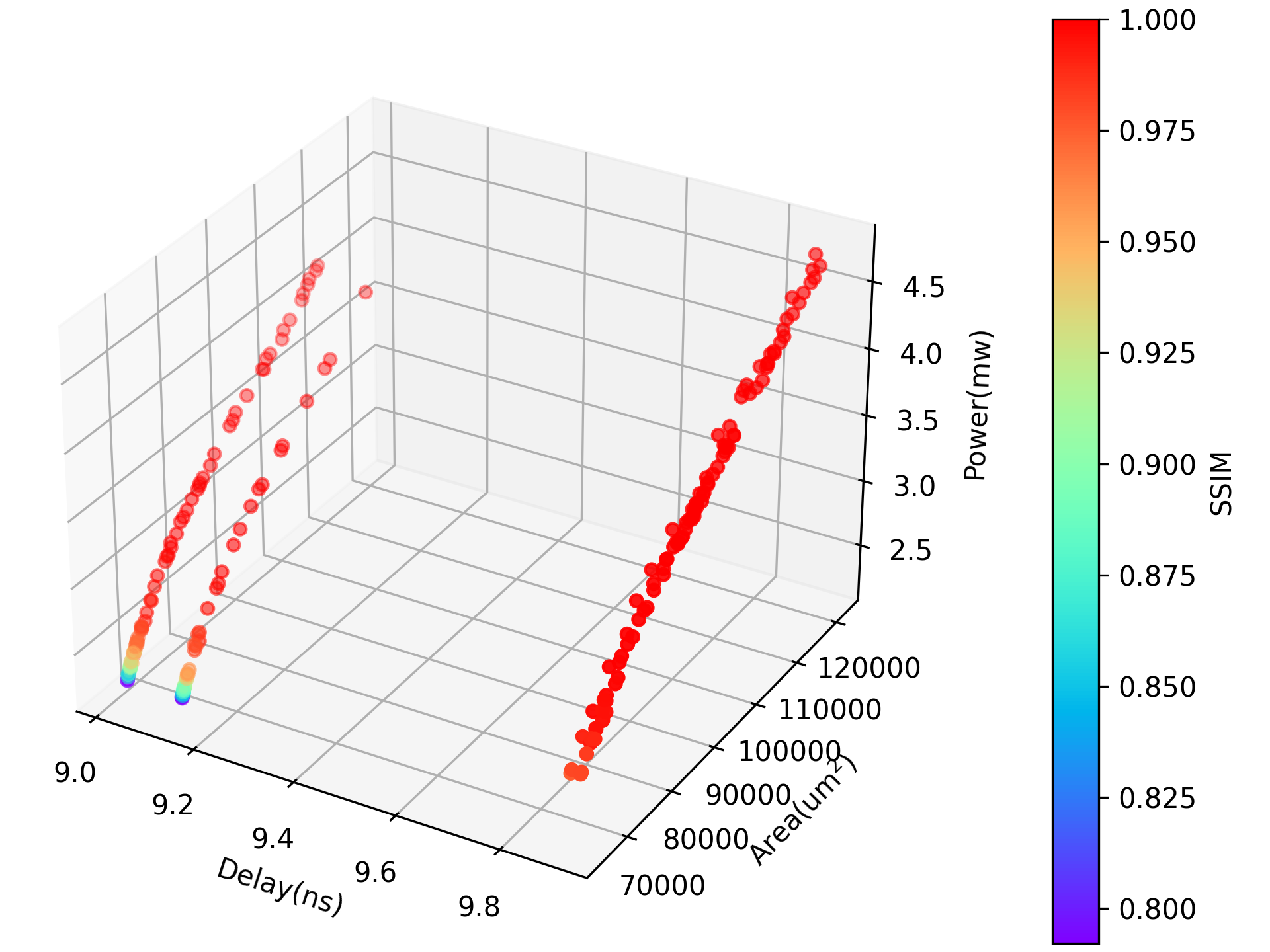} &
        \includegraphics[scale=1]{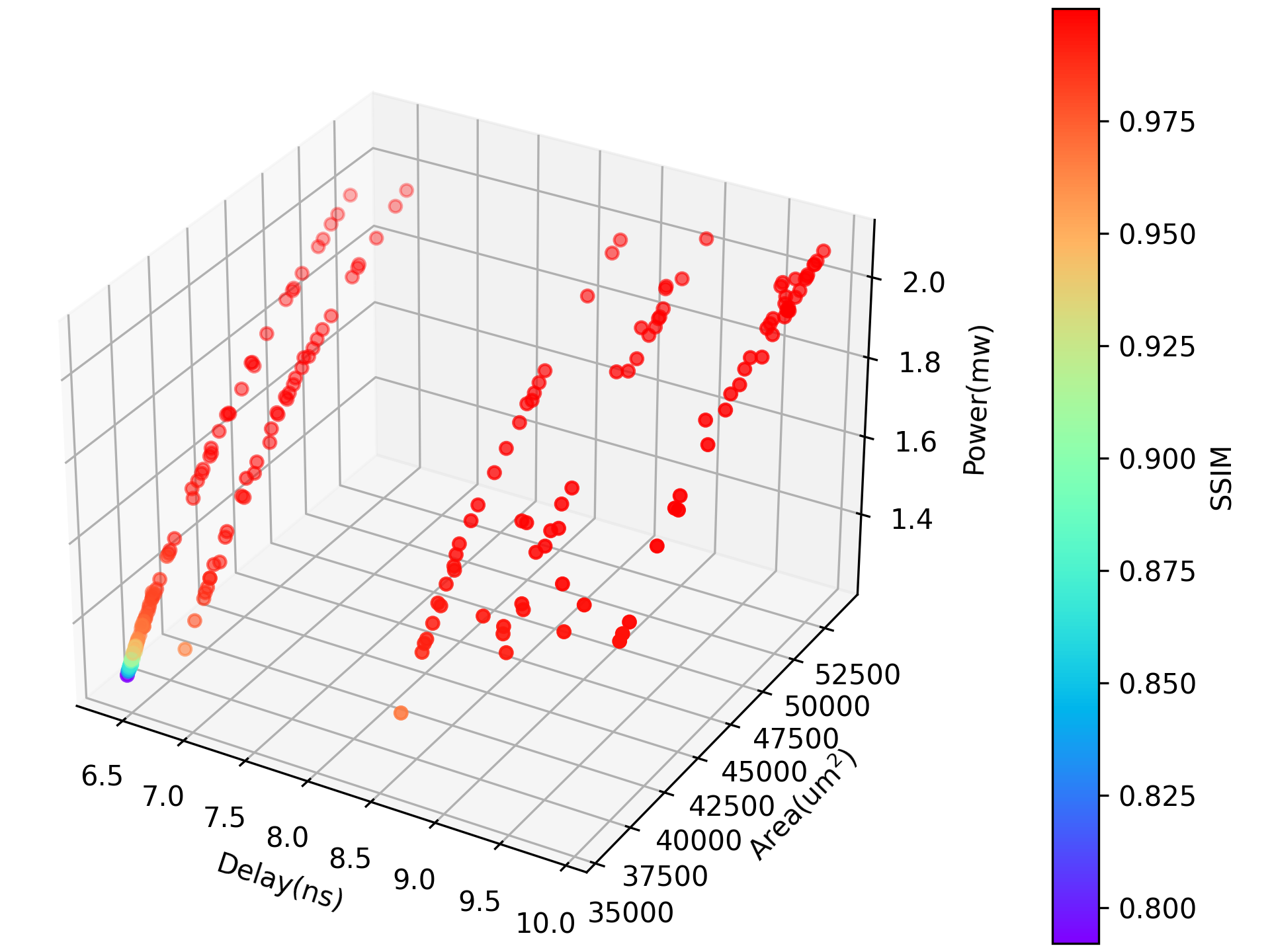} &
        \includegraphics[scale=1]{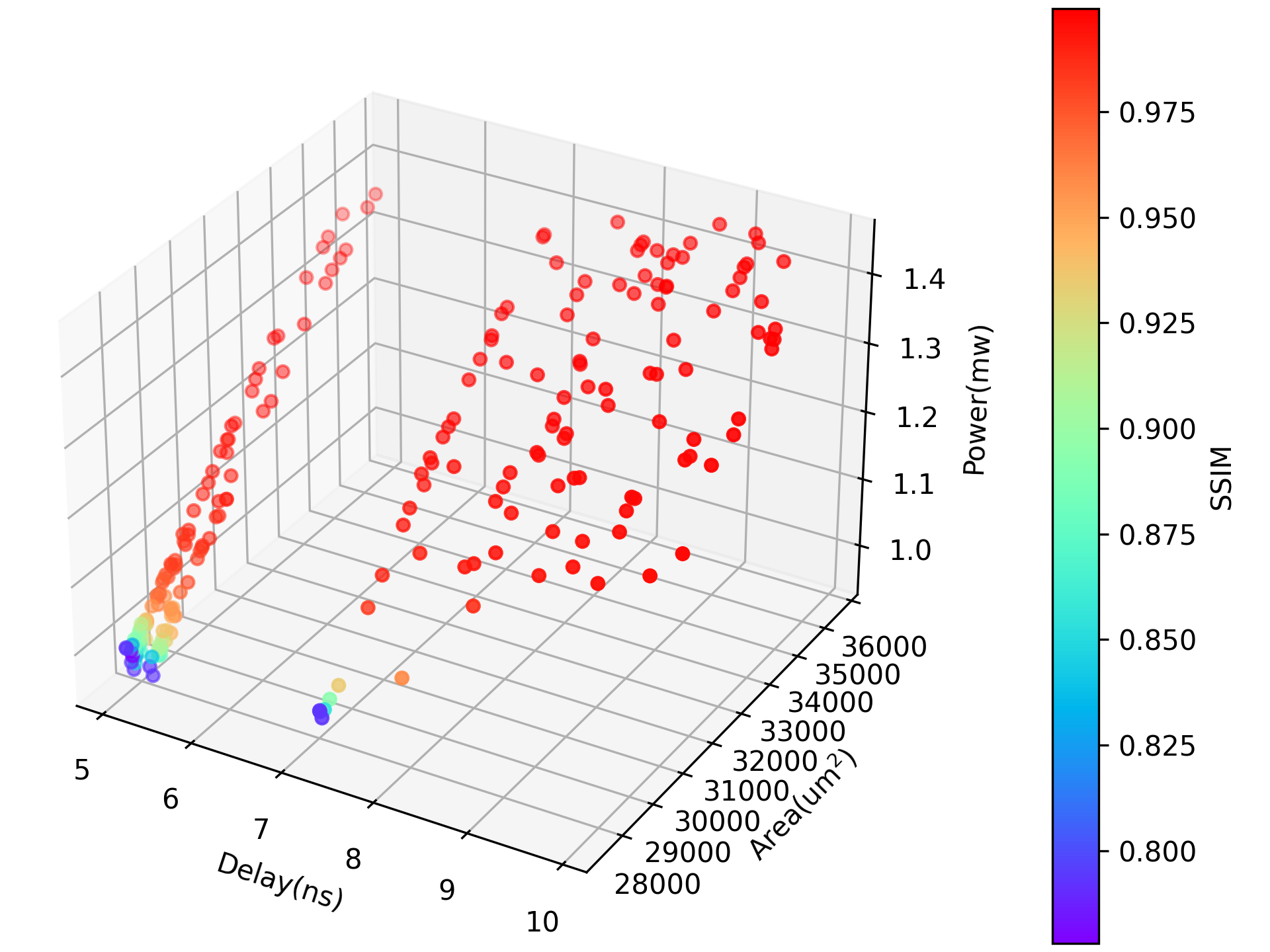}
        \\
        
    \end{tabular}
        }
\caption{Hardware Characteristics and SSIM for NSGA-II generated solutions for image processing applications.} 

\label{Fig:vsansgasolgraphs_imgproc}
\end{figure*}

Our implementation employs an enhanced version of NSGA-II that efficiently handles the increased dimensionality of the optimization problem. For each PE, the algorithm simultaneously optimizes the optimal number of LSBs to be truncated, the number of subsequent bits to be approximated, and the specific sequence of positive and negative compressors to be employed. This granular level of optimization allows each PE to be precisely tuned to its specific role within the SA, potentially leading to better overall performance and efficiency compared to approaches using static truncation and approximation parameters.
The dynamic nature of our approach introduces new considerations in error distribution analysis. With varying truncation and approximation bits across PEs, the error patterns become more complex but potentially more manageable through strategic parameter selection. This presents both challenges and opportunities in error analysis and compensation strategies, requiring more sophisticated error modeling and prediction methods. The enhanced flexibility in parameter selection offers more precise control over the accuracy-efficiency trade-off at each PE level and better adaptation to specific computational patterns within the SA.
Our research encompasses three primary floating-point~(FP) representations: IEEE 754 single precision~(FP32), NVIDIA TensorFloat-32~(TF32), and Google Brain Floating point~(BF16). While these formats share common elements - specifically a 1-bit sign and 8-bit exponent, they differ significantly in mantissa bits~(24, 10, and 7 bits respectively). This variation results in different mantissa multiplier designs: 24-bit, 11-bit, and 8-bit, corresponding to 48, 22, and 16 columns in the exact reduction stage.
By allowing each PE to operate with its own optimized set of parameters, we create a more adaptable and potentially more efficient system that can better balance computational accuracy with hardware efficiency requirements. This approach represents a significant advancement in the field of approximate computing, offering new possibilities for optimization in error-tolerant computing environments. The ability to %dynamically 
adjust truncation bits, approximation bits, and compressor sequences at the individual PE level provides unprecedented control over the accuracy-efficiency trade-off, potentially leading to more efficient and effective implementations in real-world applications.

The tables presented henceforth contains the optimized SA configurations, where the $T$-Bits column indicates the number of truncated bits in the mantissa multiplier, and the $A$-Bits column shows the number of approximated bits and compressor configuration represents the arrangement of compressors. In each column, the values are arranged in row-major order. The compressor configuration uses numbers $1-8$ to represent specific compressor configurations derived from the list of 8 as stated in ~\cite{bindu-error-diluted}. These configurations correspond to positive non-interleaving, negative non-interleaving, positive stage-wise, negative stage-wise, positive column-wise, negative column-wise, positive stage-cum-column-wise and negative stage-cum-column-wise multipliers respectively. The combination of the 3 above mentioned parameters defines the multiplier type used in a particular PE of the array, providing a compact notation for the complex arrangements of approximate multipliers. This diverse set of configurations, discovered by the NSGA-II algorithm, demonstrates various optimized solutions that balance accuracy/SSIM against hardware efficiency metrics such as delay, area, and power consumption across all three data formats (FP32, TF32, and BF16). 

\section{Experimental Results and Discussions}

% In this section, we present the experimental evaluations of our proposed floating-point multiplier designs, synthesized using an ASIC flow with the gpdk45nm technology node and implemented through the Cadence Genus tool. Our study focused on three key application domains: image processing, convolutional neural networks~(CNNs), and JPEG compression. The evaluations assessed the trade-offs between computational accuracy, resource efficiency, and hardware performance metrics, highlighting the versatility of the NSGA-II-optimized designs in real-world tasks.

The experimental evaluation framework encompassed a comprehensive synthesis and analysis methodology, incorporating both benchmark designs and solutions optimized through NSGA-II. The implementation utilized an ASIC flow based on the gpdk45nm technology node, executed via the Cadence Genus tool.
The experimental methodology focused on three distinct application domains: image processing, CNNs, and JPEG compression. Each domain presented unique computational requirements and precision constraints, enabling a thorough assessment of the proposed floating-point multiplier designs.

\begin{table*}[]
\caption{Hardware Characteristics of benchmark SOTA based Exact SAs~\cite{sota1, sota3}.} 
\label{Tab:exactsa}
\renewcommand{\arraystretch}{1.1}
\resizebox{1\textwidth}{!}{
\begin{tabular}{|c|ccc|ccc|ccc|}
\hline
SA Config & Delay (ns) & Area ($\mu$m$^2$) & Power ($\mu$W)) & Delay (ns) & Area ($\mu$m$^2$) & Power ($\mu$W)) & Delay (ns) & Area ($\mu$m$^2$) & Power ($\mu$W)) \\ \hline
          & \multicolumn{3}{c|}{FP32}                              & \multicolumn{3}{c|}{TF32}                              & \multicolumn{3}{c|}{BF16}                              \\ \hline
3$\times$3       & 19.668     & 49152.582                    & 728.519    & 14.548     & 14793.894                    & 239.253    & 11.806     & 9752.13                      & 168.419    \\
5$\times$5       & 19.398     & 148125.672                   & 2321.99    & 14.765     & 47699.082                    & 854.958    & 11.806     & 32202.72                     & 616.266    \\
8$\times$8       & 19.753     & 396673.283                   & 6333.03    & 14.715     & 130015.404                   & 2421.22    & 11.806     & 89811.252                    & 1774.04    \\ \hline
\end{tabular}
}
\end{table*}

The NSGA-II optimization framework is executed with a population size of 100 over 1000 iterations to efficiently explore the design space and obtain Pareto-optimal systolic array (SA) configurations. Since NSGA-II is a multi-objective optimization algorithm, the fitness functions consist of both application-level and hardware-level evaluation metrics. The application-level quality metric is represented by either Structural Similarity Index Measure (SSIM) or classification accuracy, depending on the target application. In addition, hardware-oriented fitness metrics include worst-case delay for performance evaluation, hardware footprint for resource utilization, and power consumption of the corresponding SA configuration.
The delay, power, and area metrics are estimated based on the configuration of each processing element (PE). For every PE configuration, the corresponding hardware metrics are extracted using the Cadence Design Systems Genus synthesis tool with the GPDK 45 nm technology library. The overall hardware characteristics of the systolic array are then obtained by aggregating the individual hardware metrics of all PEs present in the array.

In the image processing domain, we evaluated the designs through averaging and Gaussian filtering operations. These applications were selected for their dual requirements of high precision computation while maintaining resilience to minor approximation-induced errors. The evaluation metrics focused on both visual quality preservation and computational efficiency improvements.
For CNN applications, we implemented the designs within the convolution layers, which are characterized by intensive matrix multiplication operations. This integration enabled a detailed analysis of the trade-offs between model accuracy and hardware efficiency metrics, including area utilization, power consumption, and processing latency. The evaluation provided crucial insights into the viability of approximate computing in machine learning architectures.
The JPEG compression evaluation centered on the DCT stage, a computationally intensive component of the compression pipeline. Performance assessment included both image quality metrics, particularly the SSIM, and hardware efficiency parameters, allowing for a comprehensive understanding of the approximation impact.

\begin{table*}[!htp]
\caption{Hardware Characteristics and Accuracy for NSGA-II generated solutions for top accuracies of image processing applications.} 
\label{Tab:vsaconvsols}
\renewcommand{\arraystretch}{1.1}
\resizebox{1\textwidth}{!}{
\begin{tabular}{|ccccccc|}
\hline
\multicolumn{1}{|c|}{TBITS}                                    & \multicolumn{1}{c|}{ABITS}                           & \multicolumn{1}{c|}{COMBO}                           & SSIM   & Delay (ns) & Area ($\mu$m$^2$) & Power ($\mu$W) \\ \hline
\multicolumn{7}{|c|}{Averaging Filter : FP32}                                                                                                                                                                                                  \\ \hline
\multicolumn{1}{|c|}{{[}38, 37, 35, 38, 39, 36, 39, 39, 36{]}} & \multicolumn{1}{c|}{{[}0, 0, 0, 0, 1, 0, 0, 1, 2{]}} & \multicolumn{1}{c|}{{[}-, -, -, -, 2, -, -, 3, 4{]}} & 1      & 16.794     & 18503.226                    & 334.844    \\
% \multicolumn{1}{|c|}{{[}39, 37, 38, 37, 39, 35, 39, 40, 39{]}} & \multicolumn{1}{c|}{{[}1, 4, 1, 1, 1, 2, 2, 1, 2{]}} & \multicolumn{1}{c|}{{[}8, 6, 4, 4, 3, 6, 6, 4, 5{]}} & 0.9998 & 16.325     & 17730.306                    & 326.218    \\
% \multicolumn{1}{|c|}{{[}41, 39, 32, 39, 42, 41, 39, 41, 40{]}} & \multicolumn{1}{c|}{{[}1, 1, 1, 1, 0, 0, 1, 0, 0{]}} & \multicolumn{1}{c|}{{[}5, 4, 7, 8, -, -, 5, -, -{]}} & 0.9986 & 15.511     & 17342.82                     & 294.5      \\
% \multicolumn{1}{|c|}{{[}42, 38, 41, 38, 38, 38, 43, 43, 42{]}} & \multicolumn{1}{c|}{{[}0, 0, 1, 1, 1, 1, 1, 1, 0{]}} & \multicolumn{1}{c|}{{[}-, -, 8, 8, 8, 6, 6, 5, -{]}} & 0.9951 & 16.183     & 16780.23                     & 307.271    \\
\multicolumn{1}{|c|}{{[}40, 42, 41, 39, 41, 40, 42, 39, 42{]}} & \multicolumn{1}{c|}{{[}0, 0, 0, 0, 0, 0, 0, 0, 1{]}} & \multicolumn{1}{c|}{{[}-, -, -, -, -, -, -, -, 4{]}} & 0.9973 & 16.226     & 16588.71                     & 294.539    \\
\multicolumn{1}{|c|}{{[}40, 43, 41, 42, 42, 42, 42, 41, 43{]}} & \multicolumn{1}{c|}{{[}0, 0, 0, 1, 0, 0, 0, 0, 0{]}} & \multicolumn{1}{c|}{{[}-, -, -, 6, -, -, -, -, -{]}} & 0.9933 & 14.924     & 15865.722                    & 272.112    \\
\hline
\end{tabular}
}
\end{table*}

\begin{figure*}[!htp]
\centering
\resizebox{0.9\textwidth}{!}{
\setlength{\tabcolsep}{0.5pt} % Default value: 6pt
    \begin{tabular}{c c c}
     % \centering
     %\setlength{\tabcolsep}{1pt}
        \textbf{\Huge MNIST : FP32} & \textbf{\Huge MNIST : TF32} & \textbf{\Huge MNIST : BF16} \\ 
    
        \includegraphics[scale=1]{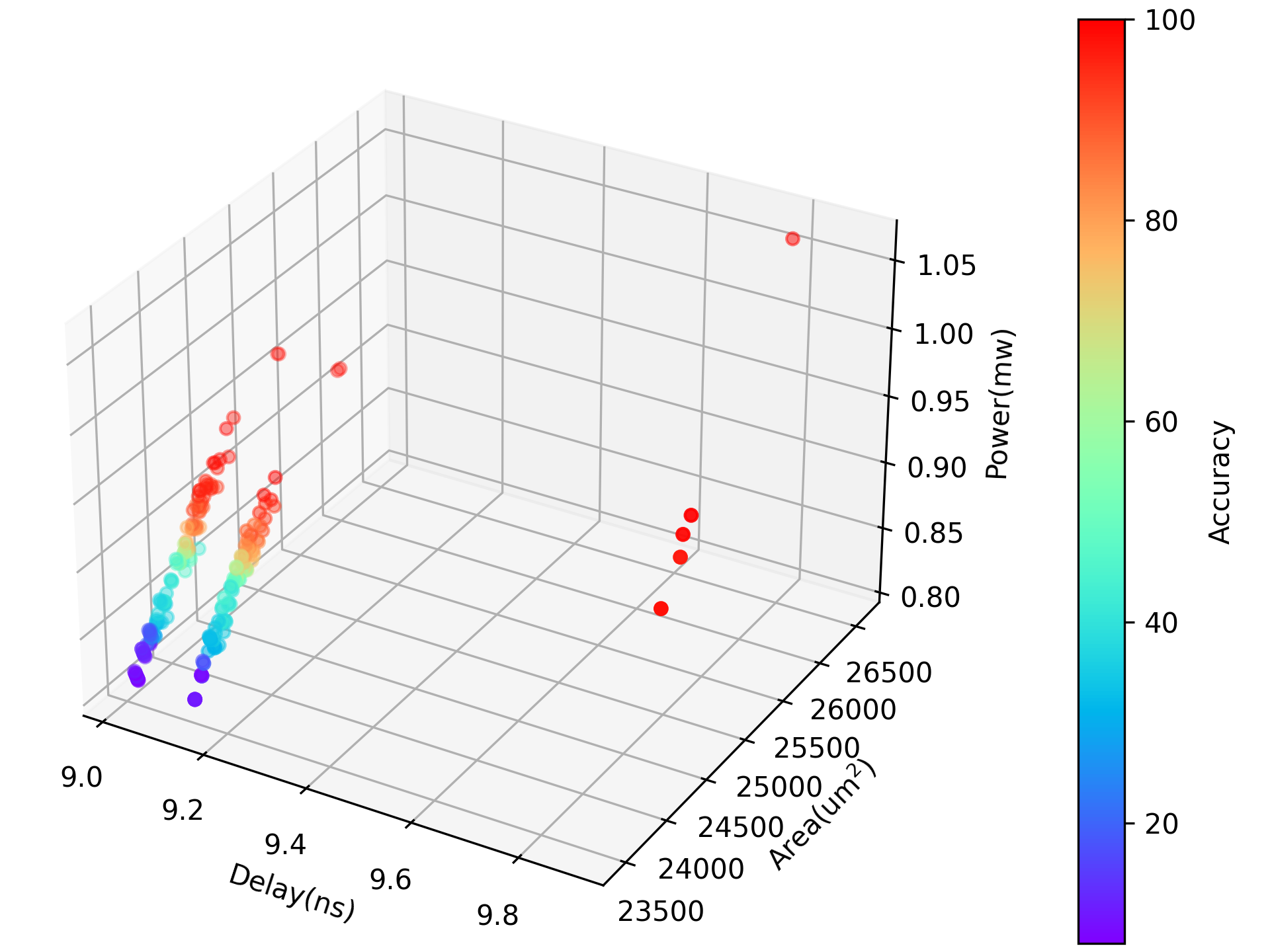} &
        \includegraphics[scale=1]{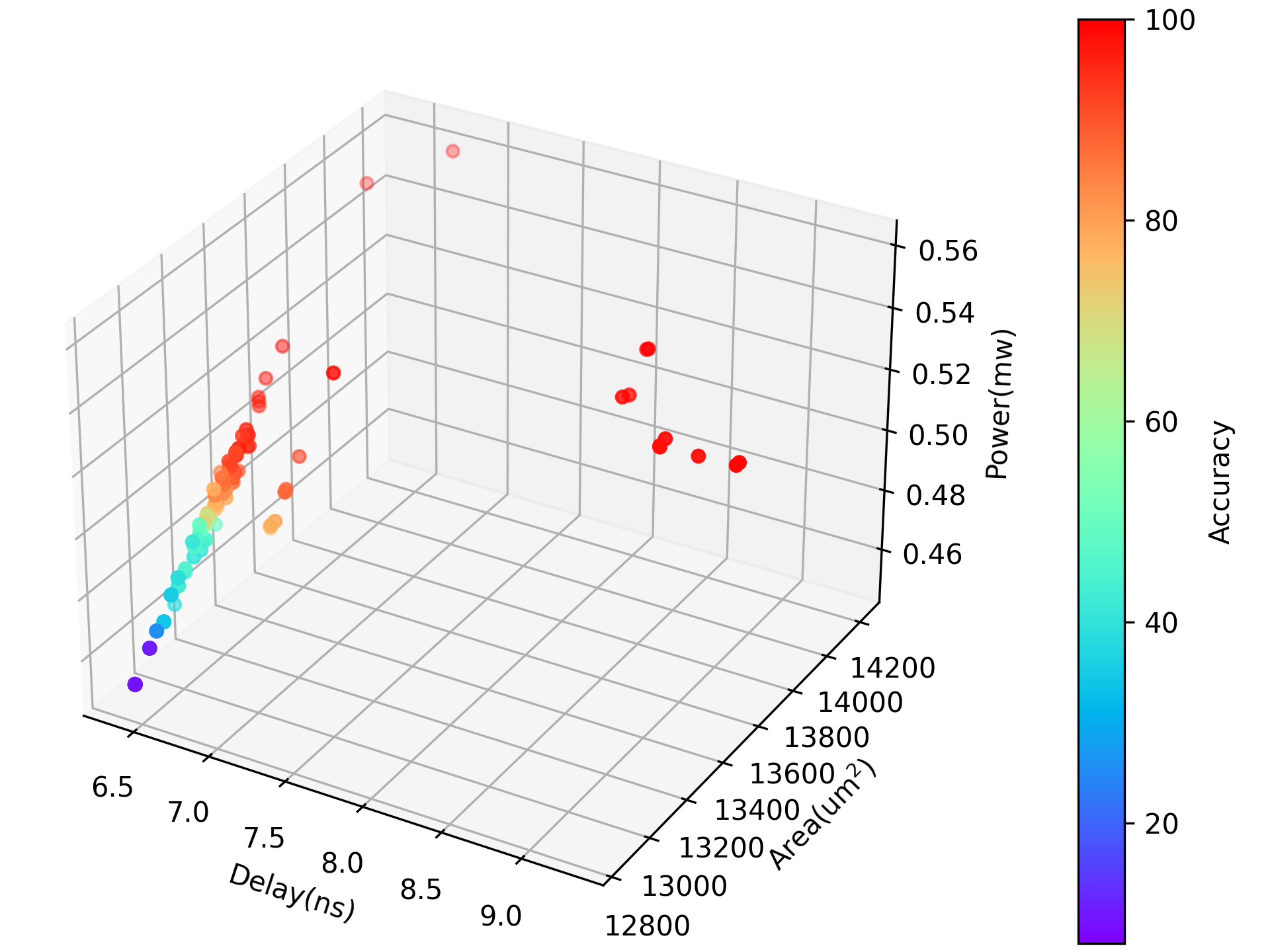} &
        \includegraphics[scale=1]{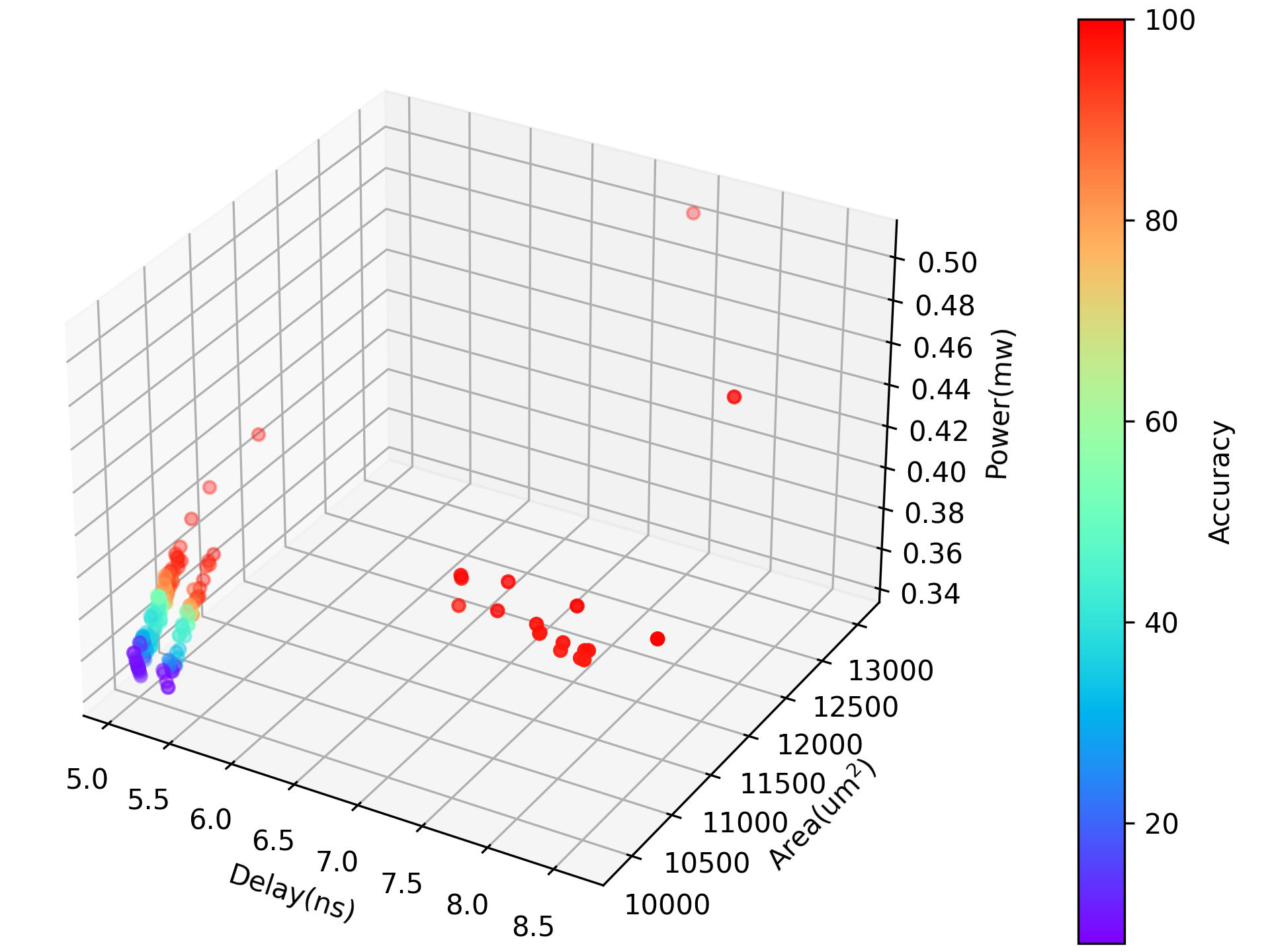}
        \\

         &  &  \\
         \textbf{\Huge FMNIST : FP32} & \textbf{\Huge FMNIST : TF32} & \textbf{\Huge FMNIST : BF16} \\ 
    
        \includegraphics[scale=1]{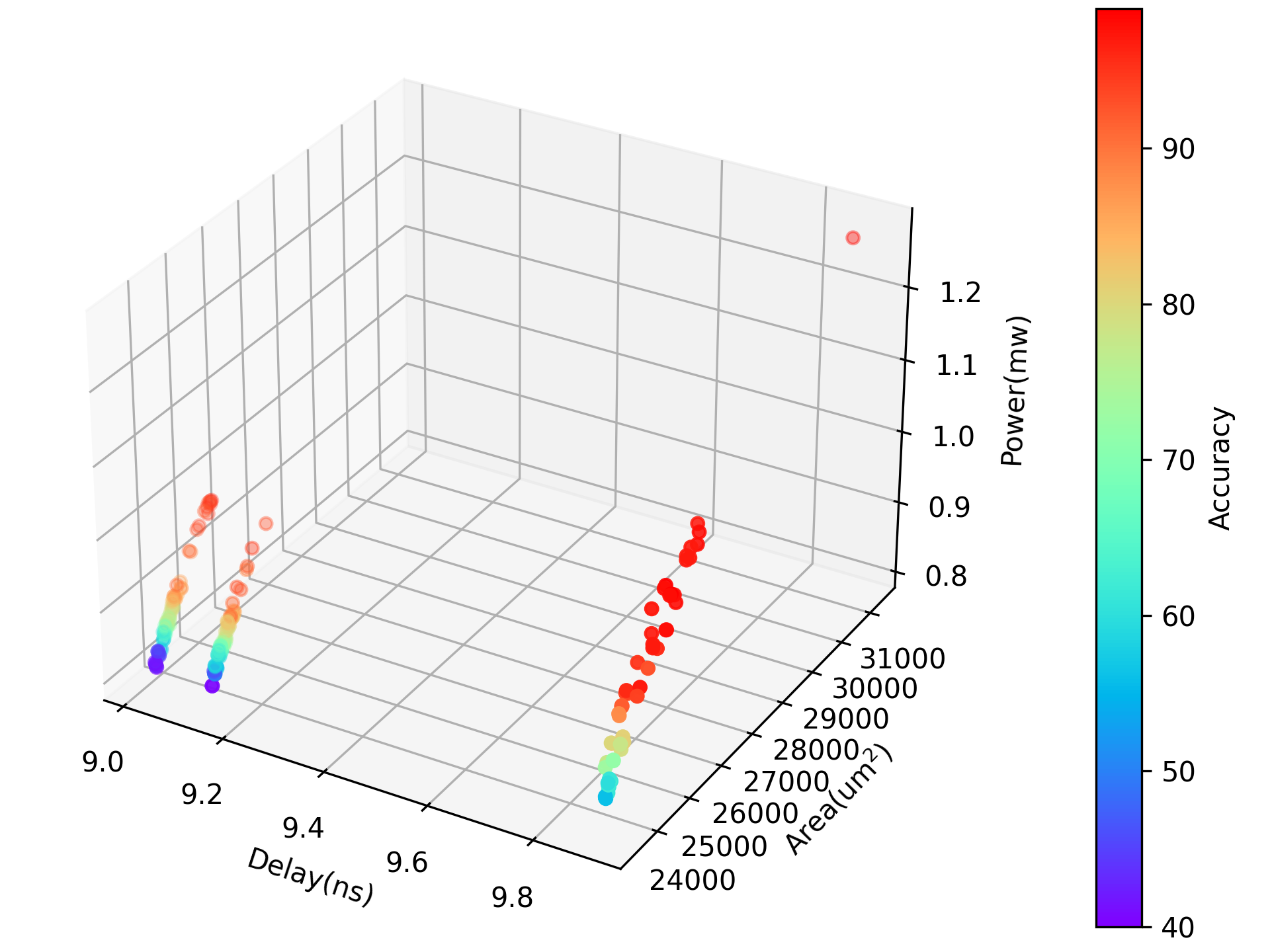} &
        \includegraphics[scale=1]{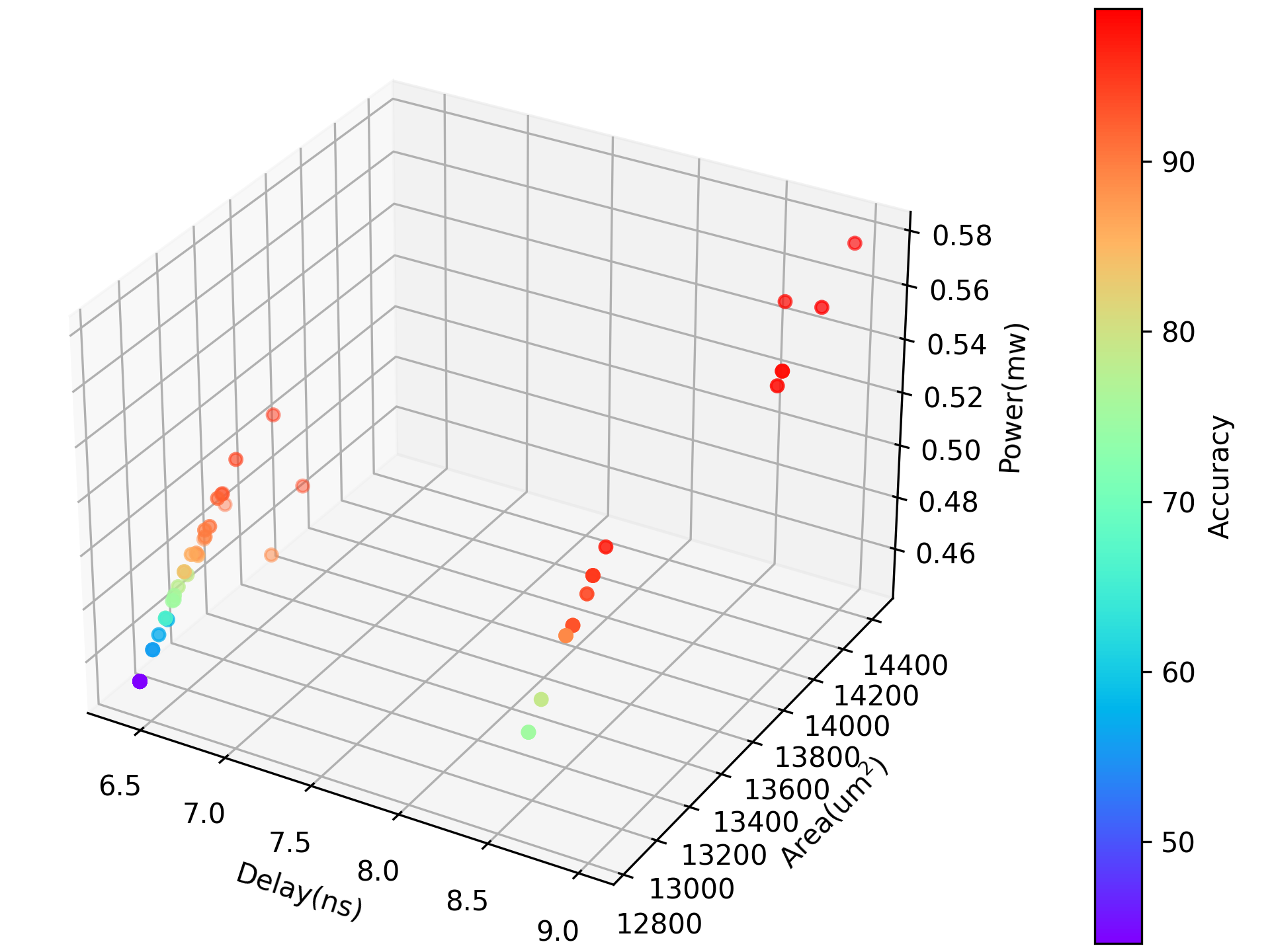} &
        \includegraphics[scale=1]{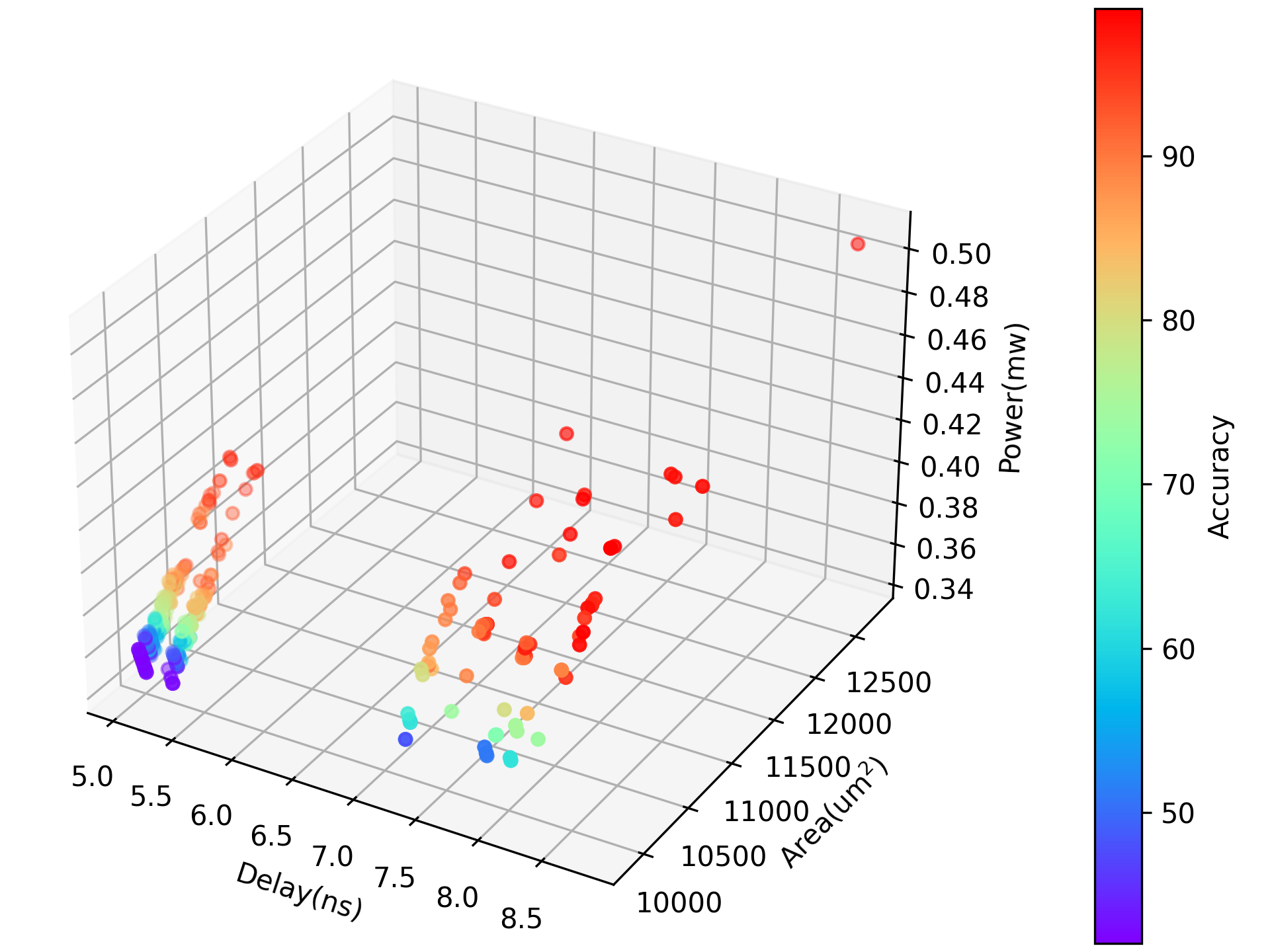}
        \\

         &  &  \\
         \textbf{\Huge CIFAR10 : FP32} & \textbf{\Huge CIFAR10 : TF32} & \textbf{\Huge CIFAR10 : BF16} \\ 
    
        \includegraphics[scale=1]{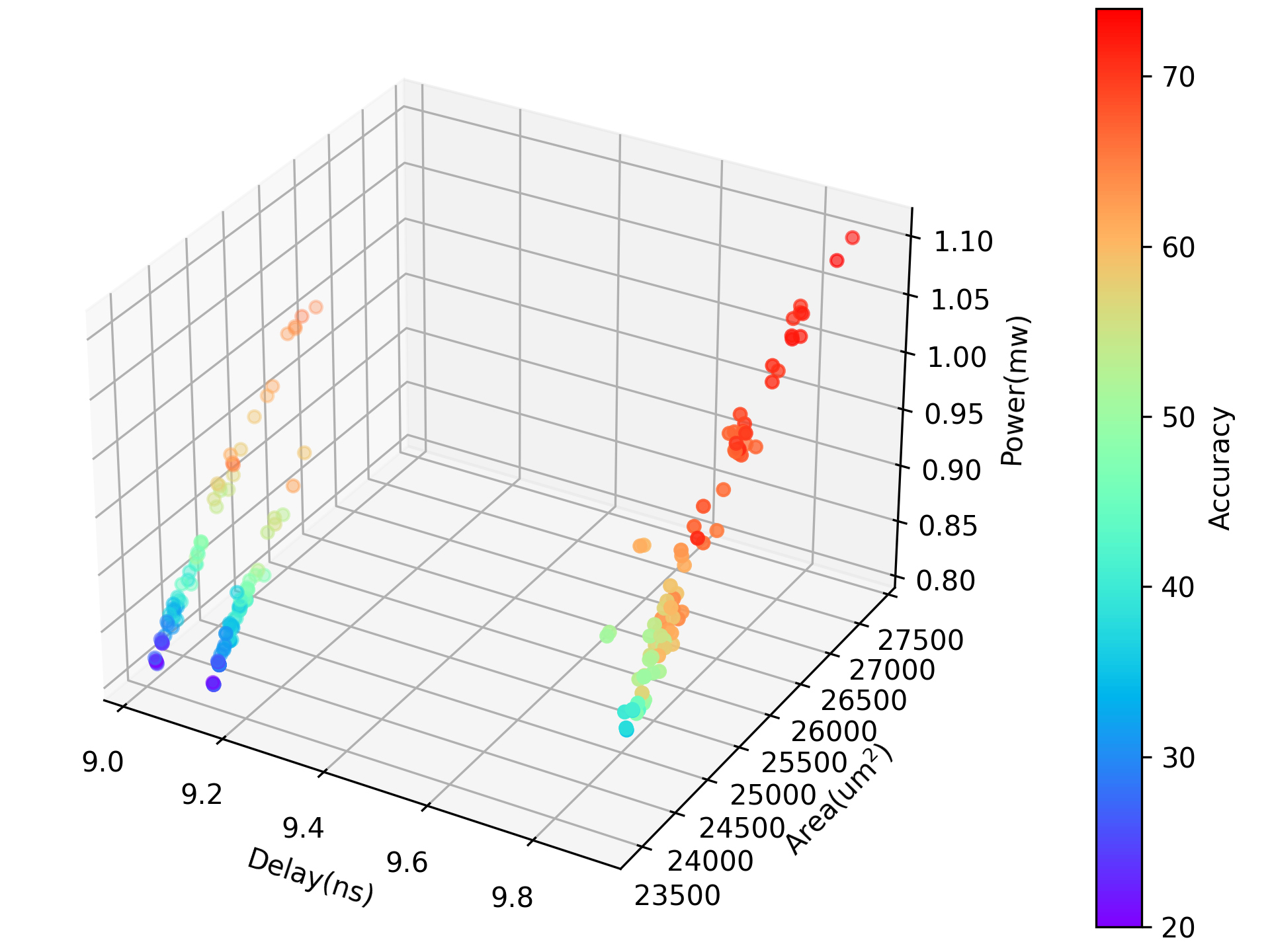} &
        \includegraphics[scale=1]{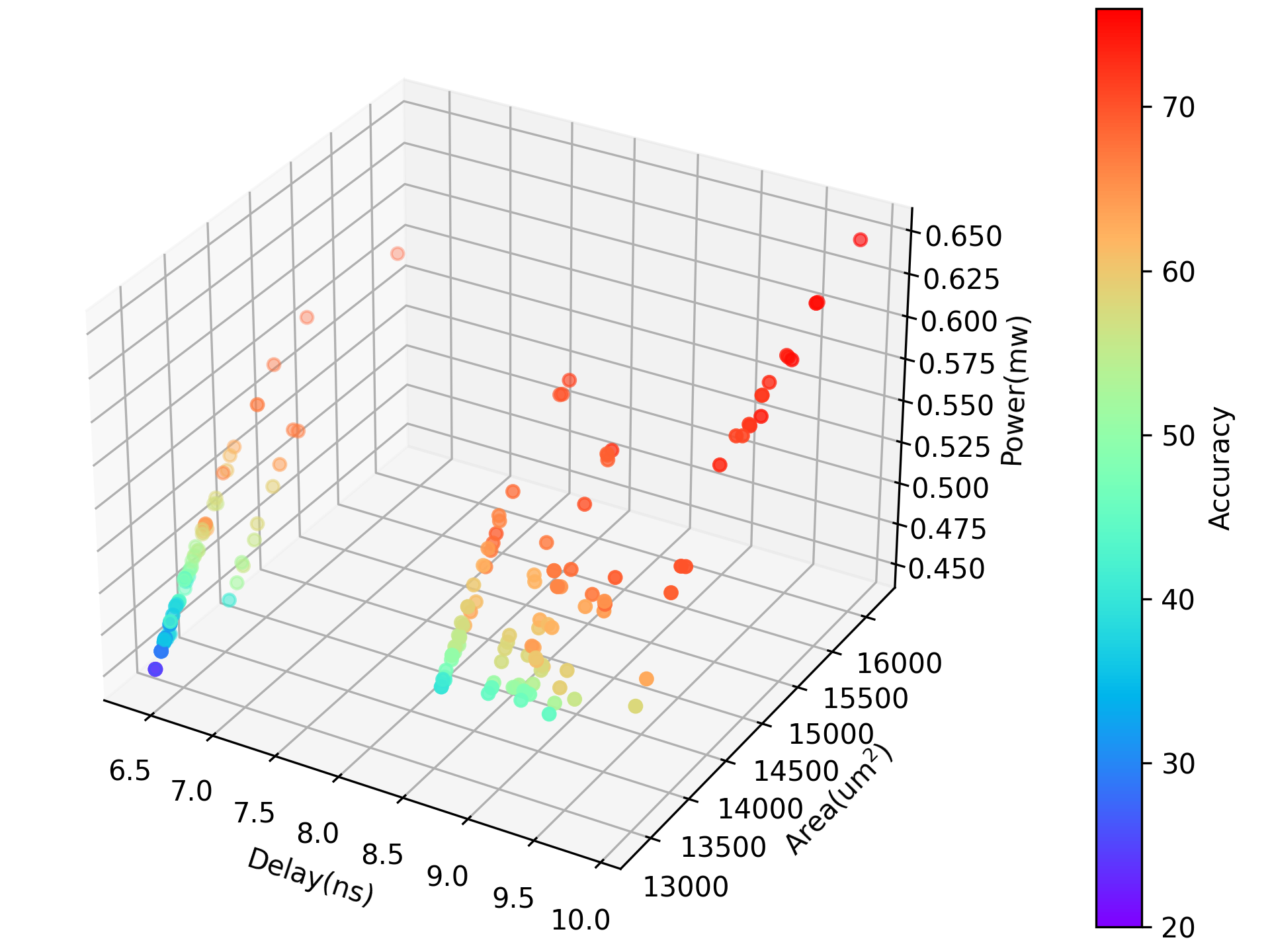} &
        \includegraphics[scale=1]{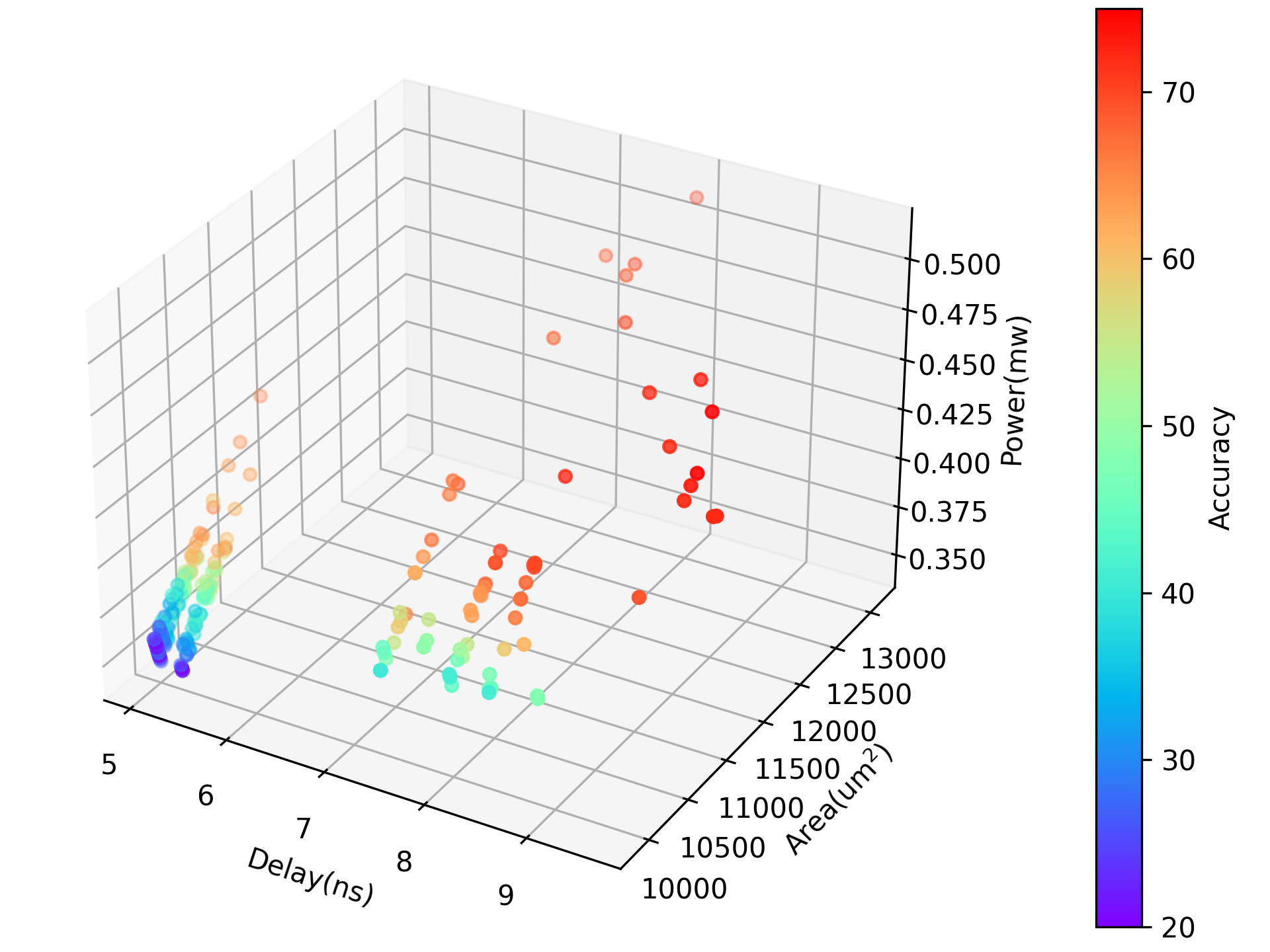}
        \\
        
    \end{tabular}
        }
\caption{Hardware Characteristics and SSIM for NSGA-II generated solutions for CNN applications.} 
\label{Fig:vsansgasolgraphs_cnnapp}
\end{figure*}

The SA configurations were tailored to each application requirements. A 3$\times$3 SA was employed for both averaging filter and CNN operations, while 5$\times$5 and 8$\times$8 architectures were implemented for Gaussian filtering and DCT operations, respectively. 
The varying SA sizes and different applications were picked to showcase the generalization of the proposed approach. 
%and varying application ensured the wide scope on the %design methodology.
The image processing and gray-scale JPEG compression evaluations utilized a standard 256$\times$256 cameraman test image, whereas color JPEG compression experiments employed a random 128$\times$128 image. The CNN implementation featured a two-layer architecture with dimensions of 32$\times$32 and 16$\times$16 for the first and second layers, respectively. Various state-of-the-art (SOTA) SA architectures~\cite{sota1,sota3} were evaluated for our study. While these architectures efficiently perform exact computations, our research focus was on approximate computing applications. To establish meaningful comparisons, we benchmarked each approximate implementation (FP32, TF32, and BF16 SA architectures) against their corresponding exact SA architecture counterparts. This approach enabled us to precisely quantify the performance-accuracy tradeoffs introduced by our approximation techniques.
The performance metrics for SA configurations using their respective exact multipliers are reported in Table~\ref{Tab:exactsa}.
The NSGA-II optimization process generated a diverse solution set, effectively balancing individual and combined performance parameters. This multi-objective optimization approach produced a comprehensive Pareto front, illustrated in subsequent figures, demonstrating the inherent trade-offs between various performance metrics. This comprehensive evaluation framework across diverse application scenarios provided robust validation of the proposed multiplier designs. The results demonstrate the effectiveness of NSGA-II-optimized designs in achieving balanced trade-offs between accuracy, computational efficiency, and resource utilization, establishing their practical viability in contemporary hardware systems.

Note that the nominal bit width of the weights in the SA is not explicitly reduced or shrunk. The weights remain stored and represented in their original floating-point formats (FP32, TF32, BF16) throughout. The approximation is applied internally within the mantissa multiplication stage of each PE during computation, not at the storage or communication level. This ensures that compatibility with standard datasets and training pipelines is preserved, and only the multiplier datapath is optimized.
Regarding interconnections within the SA, the bit width of the connections between PEs is fixed to the floating-point format under evaluation. For FP32 experiments, inter-PE connections are maintained at FP32 precision, while for TF32 and BF16, the interconnections follow their respective bitwidths. Thus, the SA datapath precision is tied to the floating-point data type being used in the experiment. Within each PE, the NSGA-II optimization selectively truncates and approximates columns in the mantissa multiplication, but the PE inputs and outputs remain compliant with the chosen floating-point standard.
This distinction clarifies that while the multiplier architectures exploit internal approximations for efficiency, the external precision of weights and interconnections remains consistent with the data type being evaluated (FP32, TF32, or BF16). This design choice ensures both functional correctness at the system level and meaningful comparison with exact implementations of the same floating-point format.

\subsection{Image Processing}

In our investigation of image processing applications, we focused on two fundamental spatial filters: a $3 \times 3$ averaging filter and a $5 \times 5$ Gaussian filter. The averaging filter, also known as a mean filter, performs simple spatial smoothing by replacing each pixel with the arithmetic mean of its neighboring pixels, making it effective for basic noise reduction. The Gaussian filter, which approximates a two-dimensional Gaussian function, provides more sophisticated smoothing while better preserving edges compared to the averaging filter, as it assigns higher weights to pixels closer to the center.
We evaluated these filters across three numerical formats: FP32, TF32, and BF16. To optimize the SA implementation, we employed the NSGA-II multi-objective optimization algorithm, which simultaneously considered four critical parameters: SSIM for output quality assessment, worst-case delay for performance evaluation, hardware footprint for resource utilization, and power consumption of the corresponding SA. Our experimental evaluation utilized the standard cameraman image as the reference benchmark.

Figure~\ref{Fig:vsansgasolgraphs_imgproc} presents the solution space obtained for CNN applications, while Table~\ref{Tab:vsaconvsols} highlights five representative configurations extracted from the aforementioned solution set for the averaging filter and gaussian filter, respectively. 
In the Table~\ref{Tab:vsacnnsols} first five rows 
mentions the top five hardware SA solutions 
and its image quality results along with hardware parameters, where each row represents a SA design. These solutions are extracted for Averaging filter of FP32 format.
The first row states $T$-bits of [38, 37, 35, 38, 39, 36, 39, 39, 36], along with 
$A$-bits of [0, 0, 0, 0, 1, 0, 0, 1, 2], with a combination~(Combo) of [-, -, -, -, 2, -, -, 3, 4], represents 9 PEs in the 3$\times$3 SA design.
The first PE is composed of 38 Truncated bits, followed by 0 Approximate bits and 
applying exact compressors to furnish product bits. Similarly the second PE is composed of 37 Truncated bits, followed by 0 Approximate bits and applying exact compressors to yield the product bits.
Similarly the last PE is composed of 36 Truncated bits, followed by 2 bits of Approximation, and $4^{th}$ (negative stage-wise) compressors to yield the product bits.

For the averaging filter implementation in FP32 format, reducing precision to a certain level resulted in an SSIM of 0.99 or higher, while achieving improvements of 26\% in critical path delay, 68\% in silicon footprint, and 62\% in power consumption. A more aggressive reduction led to an SSIM of 0.98, with corresponding enhancements of 23\%, 73\%, and 70\% across these metrics.  
The TF32 format exhibited similar behavior, where an initial level of precision reduction maintained an SSIM of 0.99, contributing to hardware gains of 22\% in critical path delay, 43\% in silicon footprint, and 35\% in power consumption. With a further decrease in precision, the SSIM remained strong at 0.98, alongside improvements of 22\% in critical path delay, 53\% in silicon footprint, and 46\% in power consumption.  All these improvements in \% are with respect to the SOTA implementations mentioned in~\cite{sota3} designs. 
For the BF16 implementation, reducing precision within a certain range preserved an SSIM of 0.99 while enhancing performance by 10\% in critical path delay, 28\% in silicon footprint, and 22\% in power consumption. At a slightly more reduced precision, the SSIM dropped to 0.98, showcasing
%still delivered significant 
hardware gains of 11\%, 44\%, and 42\% across the same metrics when compared with the exact implementations stated in~\cite{sota1}.

Similarly, for the Gaussian filter implementation in FP32 format, truncation to a specific bit width maintained an SSIM of 0.99 or greater, while providing improvements of 20\% in critical path delay, 73\% in silicon footprint, and 68\% in power consumption. Further relaxation in precision %reduction 
resulted in an SSIM of 0.98, yielding enhancements of 24\%, 78\%, and 75\% across these metrics.  
The TF32 format followed a comparable trend, where an initial reduction in precision upheld an SSIM of 0.99, leading to efficiency gains of 22\% in critical path delay, 62\% in silicon footprint, and 62\% in power consumption. With additional precision reduction, the SSIM remained at 0.98 while achieving further improvements of 28\% in critical path delay, 70\% in silicon footprint, and 69\% in power consumption.  
For the BF16 implementation, a certain level of precision reduction retained an SSIM of 0.99 while  highlighting the hardware gain by 6\% along critical path delay, 40\% along silicon footprint, and 40\% along power consumption. A further decrease in precision led to an SSIM of 0.98, with offering substantial gains of 11\%, 44\%, and 42\% across the hardware metrics. All the improvements mentioned in \% are with respect to the exact implementations discussed in~\cite{sota1, sota3}.

\subsection{Convolutional Neural Networks}

In our investigation of CNN applications, we evaluated performance across three standard benchmarking datasets: MNIST, Fashion-MNIST~(FMNIST), and CIFAR-10. 
The methodology of Precision-Aware Variable bit PEs for SA designs can be similarly adopted for different datasets and CNN layers. Three such datasets are evaluated in this work.
The implemented CNN architecture consists of multiple specialized layers operating in sequence. The first convolutional layer employs two output channels for initial feature extraction through learned kernels, detecting local patterns in the input data. This is followed by a batch normalization layer that stabilizes the learning process by normalizing activations and reducing internal co-variate shift. A max pooling layer then performs spatial dimension reduction while preserving dominant features.
The second convolutional layer expands to four output channels, enabling more complex feature detection, and is similarly followed by a batch normalization layer for training stability and a max pooling layer for further dimensionality reduction. The architecture culminates in two fully connected linear layers that integrate the extracted features for final classification.
To optimize the SA implementation of this network, we employed the NSGA-II multi-objective optimization algorithm. This algorithm simultaneously considered four critical parameters:
Network accuracy for performance validation, 
Worst-case delay for computational efficiency assessment, 
Hardware footprint for resource utilization evaluation, and
Power consumption of the corresponding SA implementation

Our methodology employed a comprehensive training and evaluation approach. The network training utilized the complete training dataset with exact multiplication operations to ensure optimal weight optimization. For inference evaluation, we selected a random subset of 100 images as our test set. While the training phase employed exact multiplication, the inference phase incorporated our proposed approximate multipliers, introducing opportunities for hardware optimization. The inference accuracy obtained using these approximate multipliers served as a key optimization parameter in the NSGA-II algorithm, enabling us to evaluate the practical trade-off between computational efficiency and model performance. This approach allowed us to assess the impact of approximation techniques on real-world network performance while maintaining computational feasibility in our optimization process.

\begin{table*}[]
\caption{Hardware Characteristics and Accuracy for NSGA-II generated solutions for top accuracies for CNN application.} 
\label{Tab:vsacnnsols}
\renewcommand{\arraystretch}{1.1}
\resizebox{0.95\textwidth}{!}{
\begin{tabular}{|ccccccc|}
\hline
\multicolumn{1}{|c|}{TBITS}                                    & \multicolumn{1}{c|}{ABITS}                           & \multicolumn{1}{c|}{Combo}                           & Accuracy & Delay (ns) & Area ($\mu$m$^2$) & Power ($\mu$W)) \\ \hline
\multicolumn{7}{|c|}{MNIST : FP32}                                                                                                                                                                                                           \\ \hline
% \multicolumn{1}{|c|}{{[}45, 43, 46, 43, 43, 46, 44, 43, 45{]}} & \multicolumn{1}{c|}{{[}0, 0, 0, 2, 0, 0, 1, 0, 2{]}} & \multicolumn{1}{c|}{{[}-, -, -, 7, -, -, 6, -, 6{]}} & 100  & 14.197     & 8678.592                     & 142.784    \\
\multicolumn{1}{|c|}{{[}46, 43, 46, 43, 43, 46, 47, 44, 45{]}} & \multicolumn{1}{c|}{{[}1, 3, 1, 3, 2, 0, 0, 2, 2{]}} & \multicolumn{1}{c|}{{[}7, 8, 7, 8, 7, -, -, 7, 7{]}} & 99   & 13.496     & 6081.102                     & 97.2133    \\
\multicolumn{1}{|c|}{{[}46, 46, 46, 42, 42, 47, 46, 43, 43{]}} & \multicolumn{1}{c|}{{[}0, 1, 1, 1, 2, 0, 0, 0, 2{]}} & \multicolumn{1}{c|}{{[}-, 6, 7, 6, 7, -, -, -, 7{]}} & 98   & 9.563      & 3595.788                     & 50.1269    \\
\multicolumn{1}{|c|}{{[}46, 45, 46, 42, 43, 47, 47, 44, 45{]}} & \multicolumn{1}{c|}{{[}0, 1, 1, 3, 2, 0, 0, 1, 2{]}} & \multicolumn{1}{c|}{{[}-, 8, 7, 7, 7, -, -, 8, 7{]}} & 97   & 13.178     & 5931.648                     & 90.5751    \\
\multicolumn{1}{|c|}{{[}47, 44, 47, 42, 43, 46, 46, 44, 45{]}} & \multicolumn{1}{c|}{{[}0, 2, 0, 3, 2, 1, 0, 2, 2{]}} & \multicolumn{1}{c|}{{[}-, 6, -, 7, 7, 7, -, 7, 7{]}} & 96   & 13.55      & 5951.826                     & 91.9252    \\
\multicolumn{1}{|c|}{{[}46, 47, 46, 42, 43, 46, 46, 44, 44{]}} & \multicolumn{1}{c|}{{[}0, 0, 1, 3, 2, 0, 0, 1, 2{]}} & \multicolumn{1}{c|}{{[}-, -, 8, 7, 6, -, -, 7, 7{]}} & 95   & 10.807     & 3556.458                     & 46.09      \\
\hline

\end{tabular}
}
\end{table*}

Our experimental evaluation yielded significant findings across three benchmark datasets, demonstrating the interplay between the approximate multipliers and both inference accuracy and hardware efficiency. Figure~\ref{Fig:vsansgasolgraphs_cnnapp} presents the solution space obtained for CNN model, while Table~\ref{Tab:vsacnnsols} highlights ten representative configurations extracted from the aforementioned solution set for the various datasets mentioned.

\begin{figure*}[!htp]
\centering
\resizebox{0.9\textwidth}{!}{
\setlength{\tabcolsep}{0.5pt} % Default value: 6pt
    \begin{tabular}{c c c}
     % \centering
     %\setlength{\tabcolsep}{1pt}
        \textbf{\Huge Gray-scale Image : FP32} & \textbf{\Huge Gray-scale Image : TF32} & \textbf{\Huge Gray-scale Image : BF16} \\ 
    
        \includegraphics[scale=1]{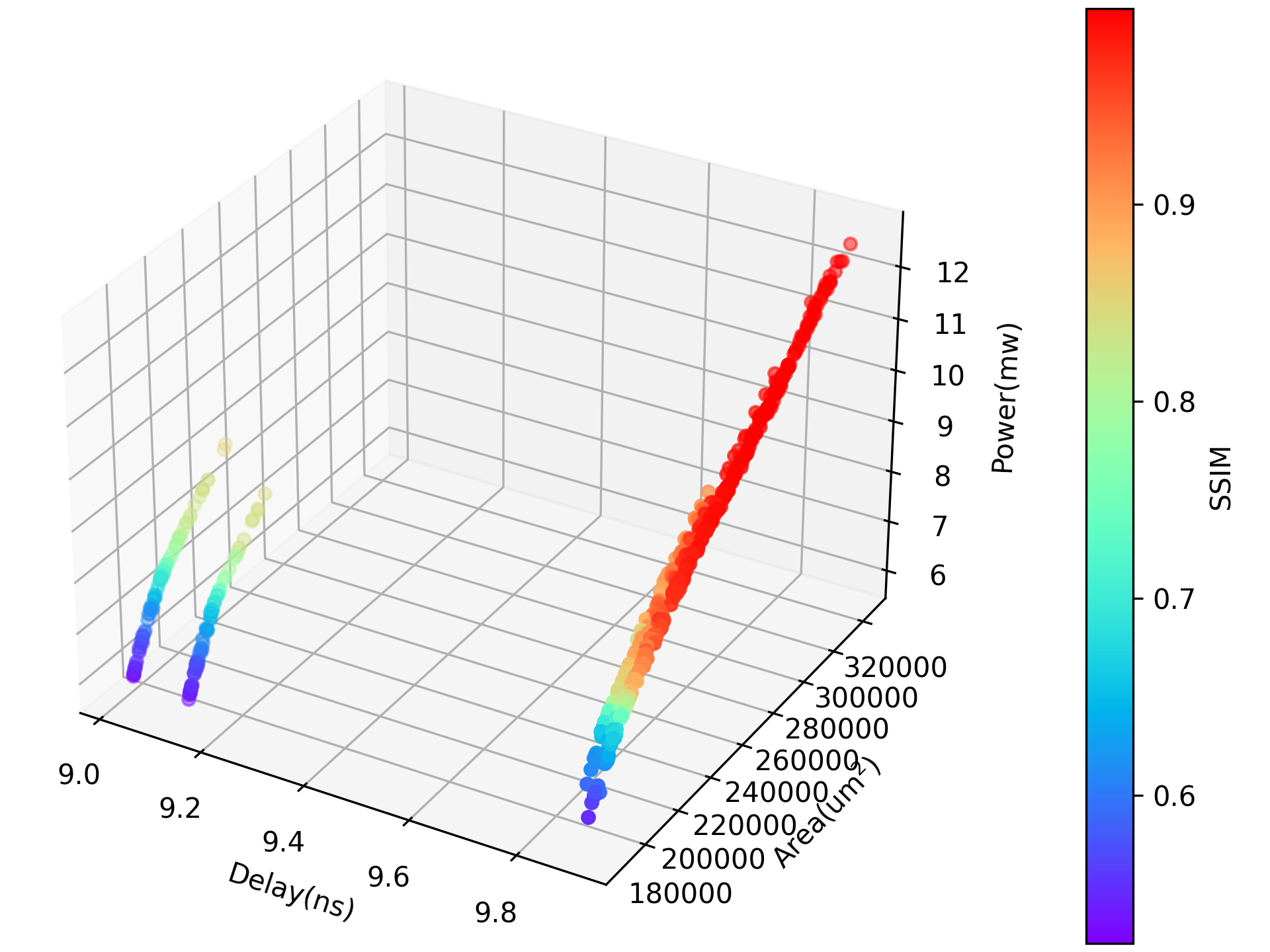} &
        \includegraphics[scale=1]{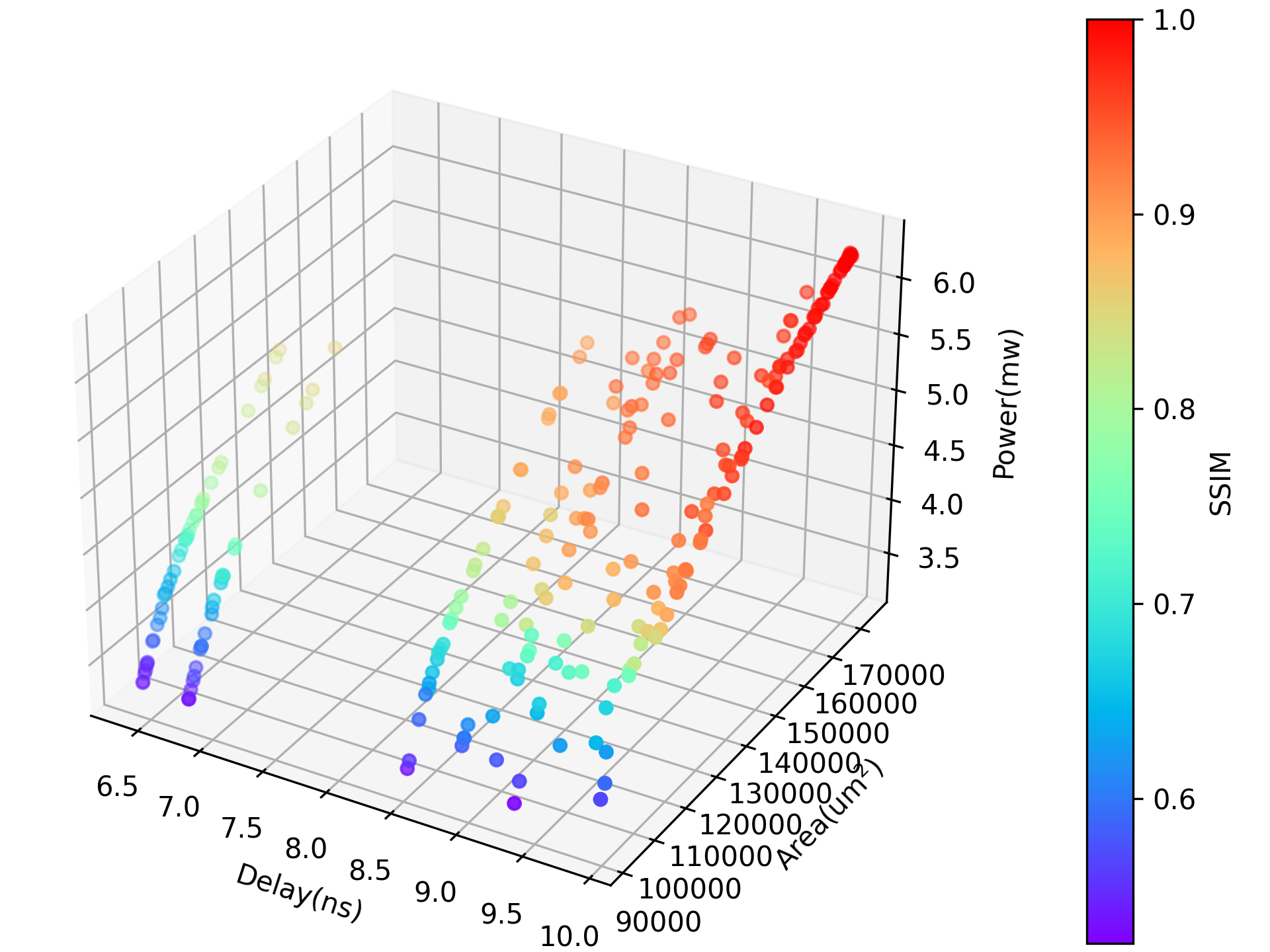} &
        \includegraphics[scale=1]{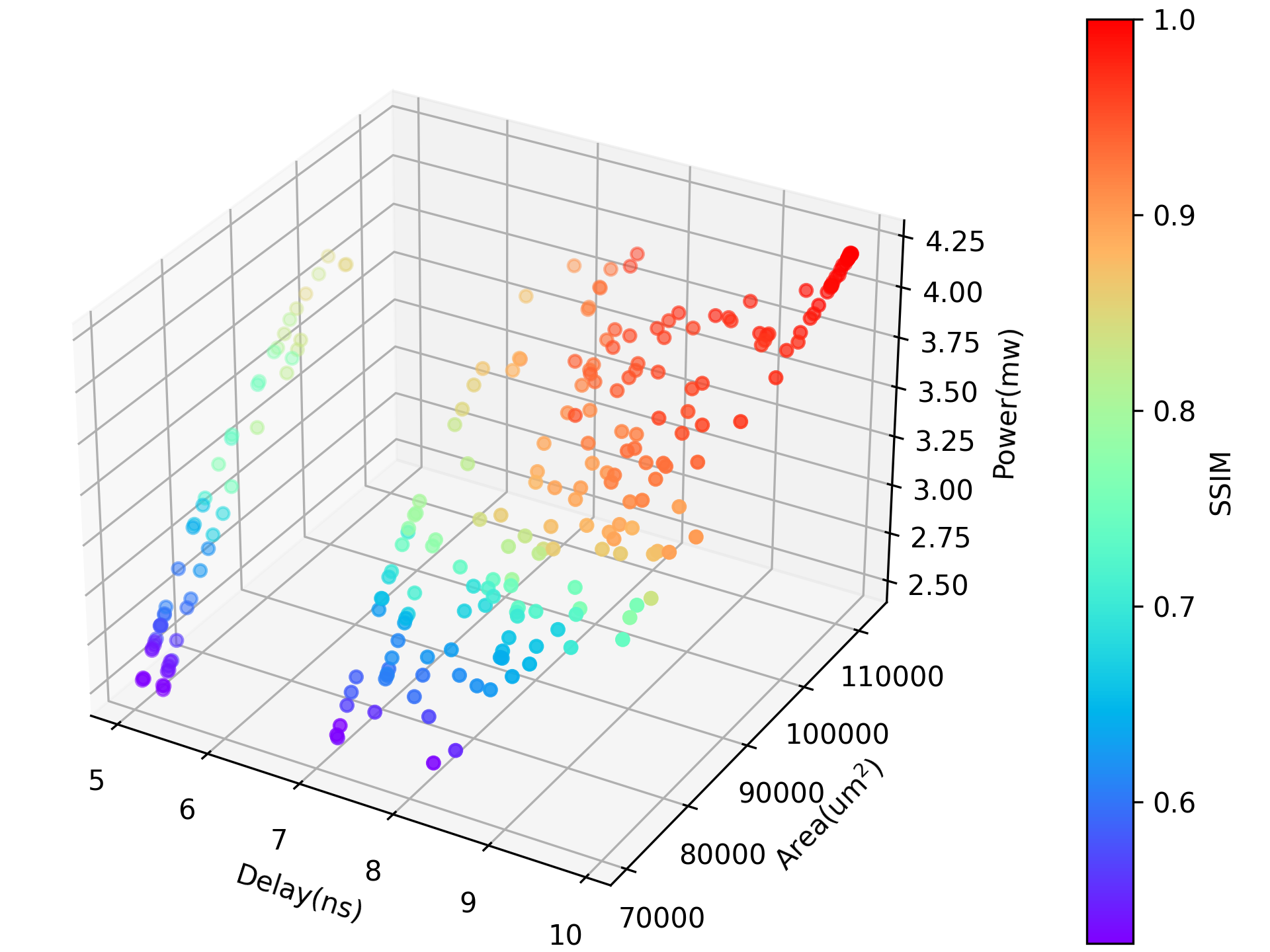}
        \\

         &  &  \\
         \textbf{\Huge Color Image : FP32} & \textbf{\Huge Color Image : TF32} & \textbf{\Huge Color Image : BF16} \\ 
    
        \includegraphics[scale=1]{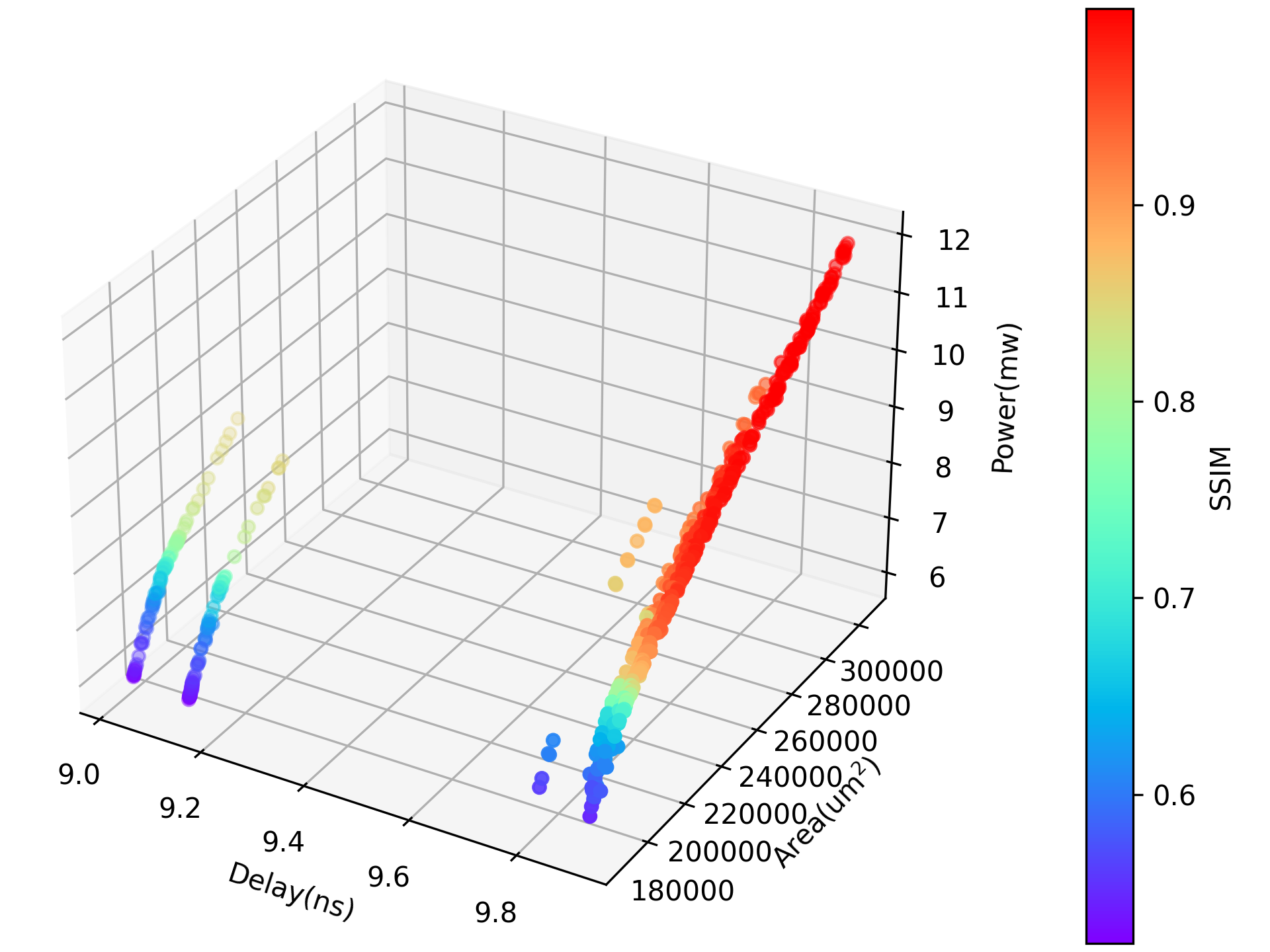} &
        \includegraphics[scale=1]{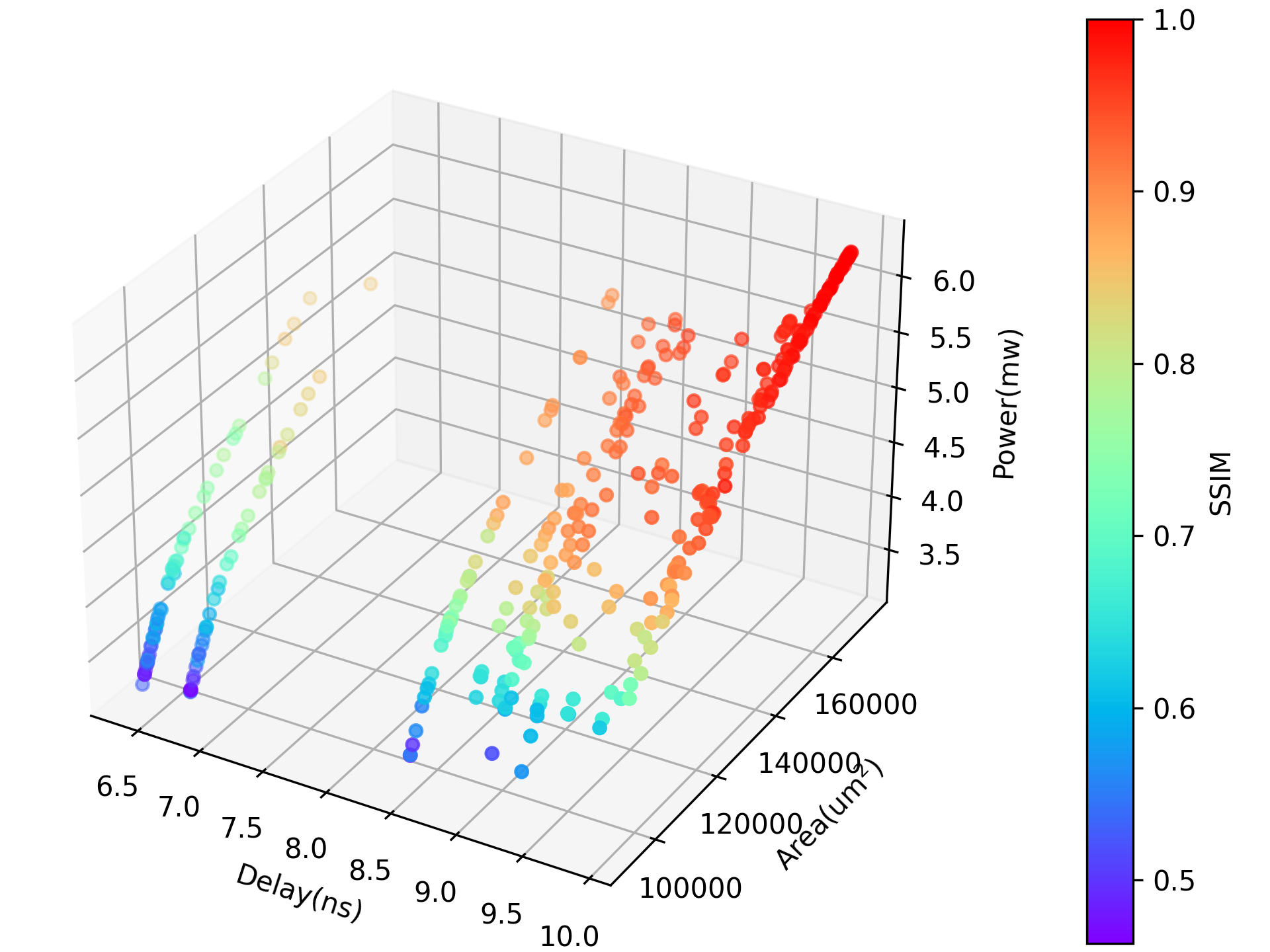} &
        \includegraphics[scale=1]{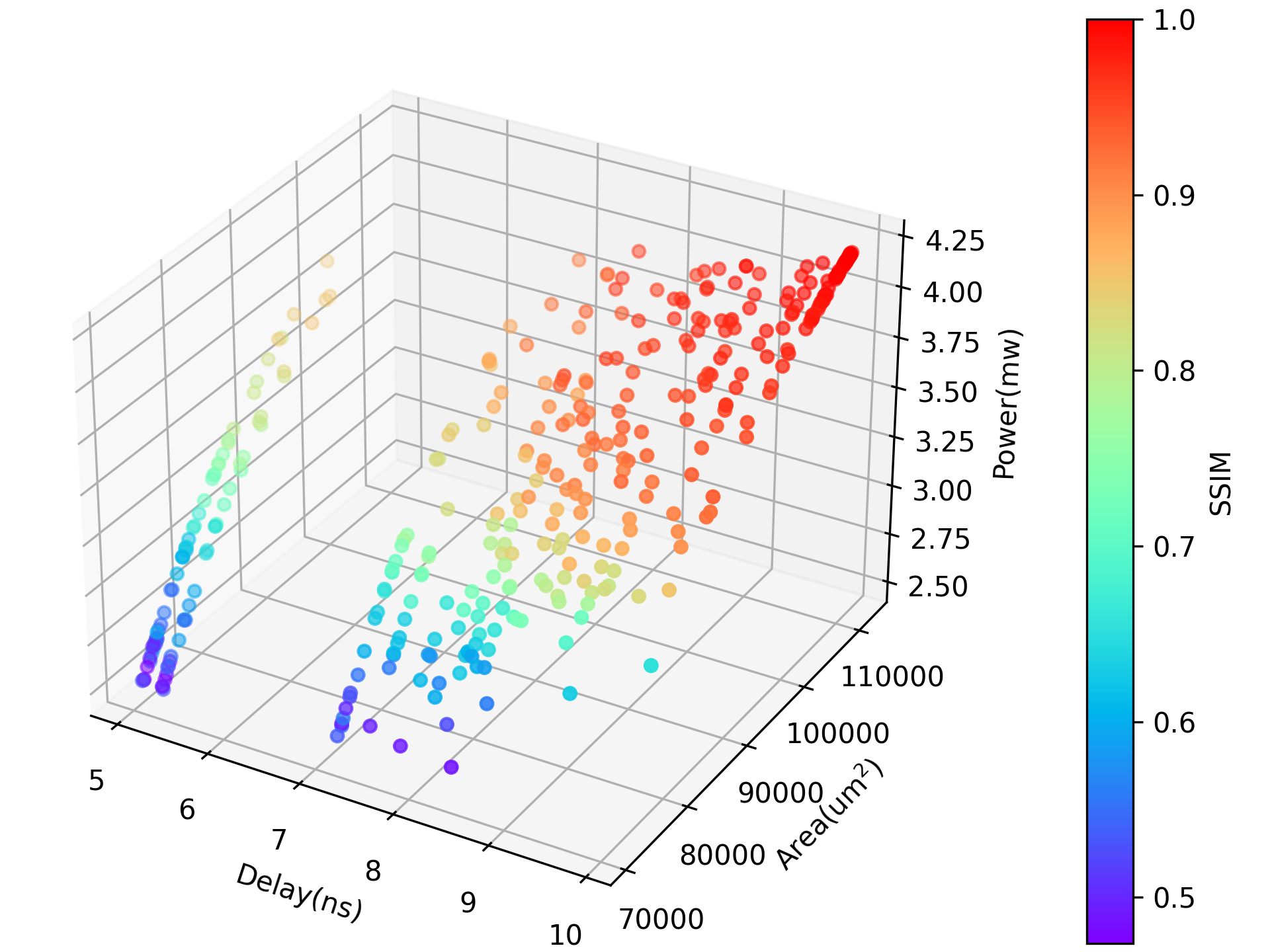}
        \\
        
    \end{tabular}
        }
\caption{Hardware Characteristics and SSIM for NSGA-II generated solutions for JPEG Compression application.} 
\label{Fig:vsansgasolgraphs_jpegcomp}
\end{figure*}

For the MNIST dataset, our baseline model achieved 99\% accuracy using conventional exact multipliers. Implementation of approximate multipliers across various numerical formats offered competitive results. The overall accuracy fluctuates between 100\% to 90\% in all the number format implementations. The FP32 implementation yields a substantial hardware gains: 27\% - 60\% reduction in critical path delay, 82\% - 92\% decrease in silicon footprint, and 80\% - 93\% improvement in power consumption. The TF32 format implementation offered hardware improvements of 31\% - 68\%, 59\% - 83\%, and 56\% - 85\% across the aforementioned metrics. Similarly, the BF16 format demonstrated hardware gains 
%efficiency gains 
of 22\% - 46\%, 63\% - 78\%, and 64\% - 82\% across the same metrics.

The Fashion-MNIST dataset yielded particularly noteworthy results. From a baseline accuracy of 97\% with exact multipliers, our approximate sequence of multipliers in the PE slots 
%implementations 
demonstrated substantial gains including performance. The accuracy that was achieved across all the number formats is 99\% to 90\%. The FP32 format achieved hardware efficiency improvements of 24\% - 31\%, 74\% - 83\%, and 71\% - 82\% in critical path delay, silicon footprint and power consumption respectively. The TF32 implementation maintained comparable accuracy ranges while achieving improvements of 24\% - 35\%, 55\% - 67\%, and 50\% - 66\% across the aforementioned metrics. The BF16 format exhibited similar performance characteristics, with efficiency gains of 16\% - 25\%, 52\% - 60\%, and 52\% - 60\% across the same metrics.

Beginning from a baseline accuracy of 68\% with exact multipliers for CIFAR-10 dataset, revealed significant improvements across all implementations. The FP32 format demonstrated accuracy between 74\% and 60\%, concurrent with hardware gains of 21\% - 54\%, 66\% - 92\%, and 60\% - 93\%. The TF32 achieves an accuracy of 76\% - 60\% while achieving the hardware benefits of 24\% - 61\% reduction in critical path delay, 44\% - 83\% improvement in silicon footprint and 37\% - 85\% improvement in power consumption. For the accuracy range of 75\% - 60\% BF16 achieves an improvement of 6\% - 45\%, 32\% - 73\% and 26\% - 78\% across the same hardware metrics.

A significant discovery emerged from our investigation: the implementation of approximate multipliers not only enhanced hardware efficiency but, in several instances, improved inference accuracy compared to exact multiplication implementations. This phenomenon was particularly evident in the Fashion-MNIST dataset, where accuracy increased from 97\% with exact multipliers to 99\% with approximate implementations. The CIFAR-10 dataset exhibited a similar trend, showing marked improvement from 68\% to 76\% which is greater than 74\% which is obtained from the previous work. This counter-intuitive finding suggests that approximate computing may introduce an implicit form of regularization, enhancing the network's generalization capabilities while simultaneously providing substantial hardware efficiency benefits.

\subsection{JPEG Compression}

\begin{table*}[]
\caption{Hardware Characteristics and SSIM for NSGA-II generated solutions for top accuracies in JPEG Compression.} 
\label{Tab:vsajpegsols}
\renewcommand{\arraystretch}{1.1}
\resizebox{1\textwidth}{!}{
\begin{tabular}{|ccccccc|}
\hline
\multicolumn{1}{|c|}{Tbits}                                                              & \multicolumn{1}{c|}{ABITS}                                              & \multicolumn{1}{c|}{Combo}                                              & SSIM                    & Delay (ns)              & Area (um\textasciicircum{}2) & Power (mW)               \\ \hline
\multicolumn{7}{|c|}{gray-scale Image : FP32}                                                                                                                                                                                                                                                                                                               \\ \hline
\multicolumn{1}{|c|}{{[}30, 33, 31, 31, 33, 34, 30, 32, 31, 28, 31, 34, 30, 30, 29, 36,} & \multicolumn{1}{c|}{{[}1, 2, 3, 0, 1, 1, 1, 2, 1, 1, 1, 1, 1, 2, 2, 2,} & \multicolumn{1}{c|}{{[}2, 4, 3, -, 5, 2, 3, 3, 2, 2, 3, 4, 3, 3, 2, 3,} & \multirow{4}{*}{0.9991} & \multirow{4}{*}{18.689} & \multirow{4}{*}{208641.888}  & \multirow{4}{*}{4138.73} \\
\multicolumn{1}{|c|}{36, 32, 33, 33, 33, 32, 32, 27, 26, 31, 27, 33, 32, 31, 33, 28,}    & \multicolumn{1}{c|}{1, 2, 1, 2, 0, 3, 4, 4, 3, 1, 2, 0, 2, 2, 2, 2,}    & \multicolumn{1}{c|}{4, 1, 4, 3, -, 4, 2, 2, 2, 3, 3, -, 4, 3, 3, 3,}    &                         &                         &                              &                          \\
\multicolumn{1}{|c|}{32, 32, 30, 32, 27, 30, 28, 28, 30, 24, 34, 31, 32, 27, 27, 32,}    & \multicolumn{1}{c|}{2, 1, 3, 3, 2, 2, 2, 1, 3, 2, 1, 2, 0, 3, 0, 0,}    & \multicolumn{1}{c|}{2, 4, 4, 4, 2, 3, 4, 3, 4, 2, 3, 3, -, 1, -, -,}    &                         &                         &                              &                          \\
\hline
\end{tabular}
}
\end{table*}

Our investigation into JPEG compression focused on both gray-scale and color image processing using two test cases: a standard cameraman image for gray-scale analysis and a representative color image for RGB compression evaluation. The compression algorithm, fundamentally based on DCT, operates by converting spatial domain data into frequency components and selectively discarding less perceptible high-frequency information. The DCT operation processes 8$\times$8 pixel blocks, transforming image data into frequency coefficients for subsequent quantization and entropy coding.
We evaluated three numerical formats~(FP32, TF32, and BF16) and optimized the SA implementation using the NSGA-II multi-objective optimization algorithm. The optimization considered four key metrics: SSIM for output quality, worst-case delay for performance, hardware footprint for resource utilization, and power consumption of the SA implementation. Figure~\ref{Fig:vsansgasolgraphs_jpegcomp} presents the solution space obtained for image processing filters, while Table~\ref{Tab:vsajpegsols} highlights ten representative configurations extracted from the aforementioned solution set for the various datasets mentioned.

For the gray-scale image JPEG compression implementation in FP32 format, reducing precision to a certain level resulted in an SSIM of 0.99 or higher, while achieving improvements of 6\% in critical path delay, 53\% in silicon footprint, and 40\% in power consumption. A more aggressive reduction led to an SSIM of 0.98, with corresponding enhancements of 10\%, 56\%, and 44\% across the metrics mentioned before respectively.
The TF32 format exhibited similar behavior, where an initial level of precision reduction maintained an SSIM of 0.99, contributing to efficiency gains of 9\% in silicon footprint, and 4\% in power consumption. With a further decrease in precision, the SSIM remained strong at 0.98, alongside improvements of 10\% in silicon footprint, and 6\% in power consumption. The delay characterized for the evolved SA designs were comparable with exact multipliers delay.
%We didnt observe any improvement in terms of delay %it is comparable to the exact multipliers.
The BF16 format evolved SA designs with its hardware parameters were comparable with the corresponding exact multiplier factored SA implementations.
%For the BF16 implementation, as well we didnt %observe much improvement in terms of hardware metric %it is comparable with the exact implementation.

Similarly, implementation of JPEG compression for the color image
 in FP32 format with truncation to a specific bit width maintained an SSIM of 0.99 or greater, offering hardware gains of 5\% in critical path delay, 50\% in silicon footprint, and 39\% in power consumption. Further relaxing the precision resulted in achieving SSIM of 0.98, yielding more hardware benefits of 11\%, 57\%, and 44\% across the corresponding metrics mentioned above.
The TF32 format demonstrated a similar trend, where an initial reduction in precision preserved an SSIM of 0.99, leading to hardware gains of 5\% in silicon footprint and 3\% in power consumption. With further precision reduction, the SSIM remained robust at 0.98, yielding additional improvements of 10\% in silicon footprint and 6\% in power consumption. However, no significant improvement was observed in 
the critical path delay, and it remained comparable to exact multipliers.
In the BF16 implementation, the hardware metrics 
of the SA designs showed minimal improvements, with performance remaining largely comparable with its corresponding exact implementations.

\section{Conclusion}
This work presents a comprehensive investigation into the application of approximate floating-point multipliers within WS SAs, demonstrating their effectiveness in balancing accuracy and efficiency across diverse computational domains. The NSGA-II  framework successfully identifies optimal configurations of Processing Elements for the SA design that yield significant reductions in power consumption, silicon footprint, and computational delay while maintaining acceptable accuracy levels. Experimental results confirm that SA design for a specific image processing applications achieve up to 78\% footprint savings and 75\% power reduction without perceptible degradation in SSIM, when compared with the exact SOTA implementations. 
CNN implementations demonstrate a remarkable balance between inference accuracy and efficiency, with the CNN model trained on Fashion-MNIST achieving a peak accuracy of 99\% and CNN model trained on CIFAR-10 exhibiting improved classification performance up to 76\%. 
Notably, the findings suggest that 
the approximation-induced noise may enhance neural network generalization, offering an additional advantage beyond traditional efficiency gains. 
The FP approximated SA designs that are 
in the top 10 CNN performance
offered substantial 
hardware gains in the range of 
%82\% to 92\% 
 66\% - 92\% footprint savings, and 
 %80\% to 93\% 
 60\% - 93\% of power benefits with 
 21\% - 54\%
 %27\% to 60\% 
 improvement in the delay when compared to its corresponding exact SOTA implementations for running the CNN model trained on CIFAR-10 dataset. The TF32 and BF16 approximated SA designs also showcased substantial gains with comparable CNN accuracy.
The JPEG compression evaluation further validates the feasibility of approximate computing by preserving image quality while enhancing hardware efficiency. 
This work presents a framework to design optimal PEs configurations for precision-aware SA designs towards accelerating workloads of interest. 
%Future research will explore the integration of %these optimized approximate SAs into real-world %hardware accelerators and extend their application %to emerging AI workloads. 
The proposed approach paves way for more resource-efficient deep learning and signal processing implementations, underscoring the transformative potential of approximate computing in modern hardware design. All the designs are made freely available for easy adoption and further usage to the researchers' and designers' community.

\balance
\bibliographystyle{unsrt}
\bibliography{bib}
%\balance

\begin{comment}
\begin{IEEEbiography}[{\includegraphics[width=1in,height=1.25in,clip,keepaspectratio]{images/author-nandini.jpg}}]{Dantu Nandini Devi} 
received her Master of Science by Research Program from the International Institute of Information Technology Bangalore. She primarily worked on approximate hardware accelerators for ML applications.\\
\textbf{Email}: dantunandini.devi@iiitb.ac.in
\end{IEEEbiography}

%If you do not have or do not want to include a photo, you can use IEEEbiographynophoto as shown below:
\begin{IEEEbiography}[{\includegraphics[width=1in,height=1.25in,clip,keepaspectratio]{images/author-MadhavRao.JPG}}]{Madhav Rao} 
is a Senior IEEE Member and is working as a faculty at IIIT-Bangalore. His group is working on VLSI Architecture Design. He teaches VLSI Architecture Design and Electronics courses at IIIT-Bangalore.\\
\textbf{Email}: mr@iiitb.ac.in
\end{IEEEbiography}
\end{comment}

%\EOD
\end{document}